\documentclass[a4paper,11pt]{article}
\usepackage[top=2 cm, bottom=2 cm, left=2 cm, right=2 cm]{geometry}
\usepackage{amssymb}
\usepackage{amsmath,subfigure}
\usepackage{epsfig}
\usepackage{eepic}
\usepackage{longtable}
\usepackage{ifthen}
\usepackage{latexsym, graphicx}
\usepackage{color}
\usepackage{subfigure}

\usepackage{lineno}
\usepackage{dsfont}
\usepackage{amsfonts}
\usepackage{amsthm}
\usepackage{mathtools}
\usepackage{multirow}
\usepackage{physics}
\usepackage{graphicx}
\usepackage{caption}
\usepackage{subcaption}
\usepackage{listings}
\usepackage{relsize}
\usepackage{tikz}
\usepackage{bigints}
\usetikzlibrary{matrix,shapes,arrows,positioning,chains}

\tikzset{
    desicion/.style={
        diamond,
        draw,
        text width=3em,
        text badly centered,
        inner sep=0pt
    },
    block/.style={
        rectangle,
        draw,
        text width=10em,
        text centered,
        rounded corners
    },
    cloud/.style={
        draw,
        ellipse,
        minimum height=2em
    },
    descr/.style={
        fill=white,
        inner sep=2.5pt
    },
    connector/.style={
        -latex,
        font=\scriptsize
    },
    rectangle connector/.style={
        connector,
        to path={(\tikztostart) -- ++(#1,0pt) \tikztonodes |- (\tikztotarget) },
        pos=0.5
    },
    rectangle connector/.default=-2cm,
    straight connector/.style={
        connector,
        to path=--(\tikztotarget) \tikztonodes
    }
}

\date{}

\title{Mathematical models for the EP2 and EP4 signaling pathways and their crosstalk}
\author{Alessandra Cambi$\,^{\mathrm{a}}$, Diane S Lidke$\,^{\mathrm{b}}$, Mariya Ptashnyk$\,^{\mathrm{c}}$, Willemijn Smit$\,^{\mathrm{d}}$, Stefanie Sonner$\,^{\mathrm{d}}$ \medskip \\
\small $^{\mathrm{a}}$Department of Medical BioSciences, Radboud University Medical Center, \\
\small Nijmegen, 6525 GA, The Netherlands
\\
\small $^{\mathrm{b}}$Department of Pathology and Comprehensive Cancer Center, University of New Mexico, \\
\small School of Medicine, 87131 Albuquerque, New Mexico\\
\small $^{\mathrm{c}}$Maxwell Institute for Mathematical Sciences, Department of Mathematics,\\  \small Heriot-Watt University, Edinburgh, Scotland, UK\\
\small $^{\mathrm{d}}$ Radboud University, IMAPP - Mathematics,  Nijmegen, 6500 GL, The Netherlands}

\begin{document}

\maketitle

%

\begin{abstract}
  The G-protein coupled receptors EP2 and EP4 both transduce the signal of the lipid messenger Prostaglandin E2 (PGE2). Changes in the cell response to PGE2 can have important effects on immunity and development of diseases, however, a thorough understanding of the EP2-EP4 receptors' signaling pathways
is lacking. Experimental data show that receptor activity (indicated by cAMP expression) has a different kinetics depending on which receptor is triggered by PGE2 and that crosstalk exists between EP2 and EP4.  To better understand the underlying mechanisms and be able to predict cell responses to PGE2, we develop novel mathematical models for the cAMP signaling pathways of  EP2 and EP4 and their crosstalk. Ligand binding dynamics plays a crucial role for both, the single receptor activity and their
crosstalk.
The mathematical models can predict the qualitative cAMP levels observed experimentally and provide possible explanations for the differences and commonalities in the signaling behavior of EP2 and EP4.
As inhibition of PGE2 signaling is gaining increasing attention in tumor immunology, these mathematical models could  contribute to design effective anti-tumor therapies targeting EP2 and EP4.
\end{abstract}

 {\small \it Keyword:}
{\small G-protein coupled receptors,  cellular signaling pathways,  receptor crosstalk,  ordinary differential equations}


\section{Introduction}

Prostaglandin E2 (PGE2) is a lipid mediator that plays an important role in modulating myeloid immune cells such as macrophages and dendritic cells (DCs) through the activation of the G protein-coupled receptors (GPCRs) EP2 and EP4 \cite{Keijzer}. Both EP2 and EP4 signaling leads to an increase of intracellular cyclic adenosine monophosphate (cAMP) levels via  G$_{\alpha_s}$ protein activation. Additionally, EP4 signals through the inhibitory 
G$_{\alpha_i}$ protein that decreases cAMP production. While the key players of these signaling pathways are known, the specific contributions of EP2 and EP4 and their potential crosstalk in directing cellular outcomes are  poorly understood. We have previously shown experimentally that EP2 and EP4 signaling generate distinct cAMP production profiles  \cite{Vleeshouwers}. While EP4 induced a transient cAMP response with a sharp initial increase and then decay over time, EP2 produced a sustained cAMP response only at high PGE2 levels. Furthermore, simultaneous EP2 and EP4 activation led to reduced cAMP expression compared to the production by a single receptor with a cAMP profile similar to EP4. These data collectively suggested the possibility of a signaling crosstalk between EP2 and EP4, but the mechanisms of this crosstalk are not known.
Mathematical models specifically dedicated to EP2 and EP4 would be instrumental to simulate multiple signaling scenarios and to better predict cell responses to PGE2.

There is an extensive literature on mathematical models,  with varying level of detail and complexity, describing the G-protein activation cycles of general GPCRs, see e.g. \cite{Carvalho,Katanaev,Woodroffe}, and cAMP regulation dependent on multiple subcellular enzymes, see e.g. \cite{Saucerman, Xin}. However, there is no unifying framework to describe  the complete signaling pathways of GPCRs and most models assume that ligand receptor binding is rapid. In particular, specific models for the cAMP signaling pathways of EP2 and EP4 that can predict and explain the experimentally observed cAMP production profiles in \cite{Vleeshouwers} are lacking. 
The cAMP signaling pathways  consist of three steps. The pathway is initiated by the binding of a ligand to the receptor, the receptor then activates G-proteins downstream, and finally, activated G-proteins trigger  cAMP expression via interaction with enzymes. 
A mathematical model taking into account these three steps in the cAMP signaling cascade of the GLP-1R receptor was considered in \cite{Bridge}. Assuming that ligand-receptor binding is rapid the cAMP signaling pathway of EP2 was modeled in \cite{Leander}, but the model cannot predict the for EP2 characteristic threshold behavior in cAMP expression in response to PGE2.  

Here, we develop mathematical models that  can explain the commonalities and differences in the signaling of EP2 and EP4 and their crosstalk and can qualitatively predict experimentally observed cAMP production profiles in response to PGE2. 
The models are formulated as  systems of ordinary differential equations and describe the dynamics of the ligand PGE2 activating the receptors EP2 and EP4, the intracellular G-protein activation cycles and cAMP production dependent on  the G-proteins G$_{\alpha_s}$ and G$_{\alpha_i}$. It turns out that receptor specific ligand binding kinetics play an important role and can cause the qualitative differences in the cAMP profiles of EP2 and EP4.  The threshold behavior of EP2 in cAMP expression when varying PGE2 concentrations can be explained by higher order reactions in the binding between ligands and receptors. The temporal dynamic interactions between ligands and EP4 receptors, in particular internalization, were essential to obtain the experimentally observed sharp increase and then decrease in cAMP levels over time. Moreover, for EP4 it  is important to take competitive binding between the stimulating G$_{\alpha_s}$ and the inhibiting  G$_{\alpha_i}$ proteins into account. The sensitivity of cAMP to the positive and negative feedback from G$_{\alpha_s}$ and G$_{\alpha_i}$ proteins is also reflected in the level of cAMP when considering  crosstalk between EP2 and EP4 receptors. 

PGE2 is a very important mediator in the immune system and can induce both activating and  suppressive immune responses, for example against tumours \cite{Cuenca}. As such, targeting PGE2 and its receptors is
a promising therapeutic strategy to improve anti-tumour immune responses. To this aim, it is crucial to unravel EP2 and EP4 specific signaling behavior and to develop mathematical
models that help predicting immune cell responses in antitumour therapies targeting these receptors.

The outline of our paper is as follows: 
In Section \ref{sec:bio} we describe the three steps in the cAMP signaling cascade of EP2 and EP4 in greater detail and present the experimentally observed cAMP profiles of EP2, EP4 and their crosstalk in response to PGE2. In Section~\ref{sec:signEP2EP4} we develop mathematical models for the three steps in the cAMP signaling pathway of EP2 and EP4. 
In Section~\ref{sec:sim} we present the numerical simulation results of the models, which are then discussed in Section~\ref{sec:con}. 

\section{Biological background and experimental results}\label{sec:bio}
\subsection{Ligand binding}

GPCR signaling is initiated through the binding of a ligand to a receptor on the cell surface. The receptor becomes activated, i.e. it undergoes a conformational change and opens a binding site for G-protein complexes inside the cell \cite{Woodroffe}. Depending on the specific ligand and receptor the ligand-receptor binding kinetics can be significantly different.
EP2 and EP4 both bind to the same ligand PGE2, but the ligand-receptor dynamics shows several qualitative differences. Here, we only mention the processes relevant for EP2 and EP4.

The activation/inactivation cycle of a receptor can display zero-order, ultrasensitivity behavior
with respect to signals, i.e.\ a small change in the ligand concentration can lead to a large change in the concentration of activated G-proteins \cite{Roth}. The concentration of activated G-proteins as a function of the ligand concentration then
behaves switch-like, which resembles the responses of cooperative enzymes \cite{Markevich}, and leads to a switch-like behavior in the cAMP profile as observed for EP2, but not for EP4. 
On the other hand, when a ligand  binds to a GPCR on the cell surface the receptor can be internalized,  i.e.~some receptor molecules are internalized by the cell after ligand binding. These receptor molecules can be recycled back to the cell surface after signaling~\cite{Dey}. 
While EP4 is rapidly internalized upon ligand binding~\cite{Vleeshouwers}, EP2 is not internalized.  Another difference between EP2 and EP4 concerns the affinity of the ligand for the receptors. Radioligand binding studies indicate that the affinity of PGE2 for EP4 receptors is higher than for EP2 \cite{KONYA2013485} and the dissociation constant of EP2 and PGE2 can be an order of magnitude larger than the dissociation constant of EP4 \cite{MaJaSoMl}.

Denoting by $r$ the free receptors and by $r^*$ the receptors bound to the ligand $p$, the reactions occurring upon ligand binding are as follows
\begin{align}\label{rec_act}
 \text{for EP2 and EP4: }    r+p \longleftrightarrow r^*, \qquad \text{for EP4: } r^*  \longrightarrow \text{internalized receptor}.
\end{align}

\subsection{G-protein activation cycle}

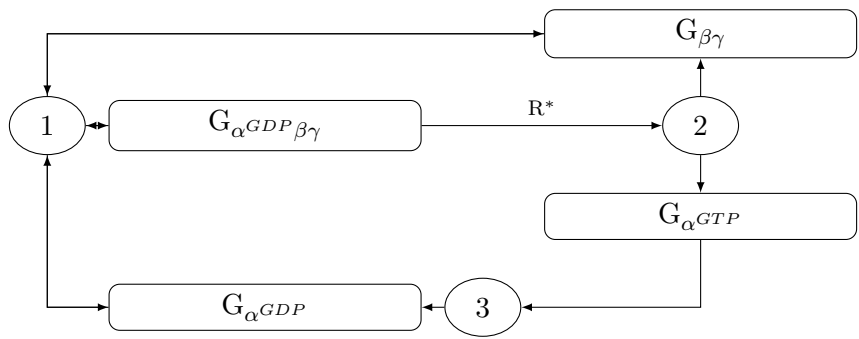
\begin{figure}[ht]
\centering

\begin{tikzpicture}
    \matrix (m)[matrix of nodes, column  sep=0.3cm,row  sep=5mm, align=center, nodes={rectangle,draw, anchor=center} ]{
                        &     &  &  |[block]| {G$_{\beta\gamma}$}                                     \\
        |[cloud]| { 1}               &    |[block]| {G$_{\alpha^{GDP}\beta\gamma}$} &    &       |[cloud]| { 2}                              \\
          & &   &    |[block]| {G$_{\alpha^{GTP}}$}                                       \\
         & |[block]| {G$_{\alpha^{GDP}}$} &     |[cloud]| { 3}           &                                             \\
    };
    \path [>=latex,->] (m-2-1) edge (m-2-2);
    \path [>=latex,->] (m-2-2) edge (m-2-1);
    \path [>=latex,->] (m-2-2) edge node[auto] {\scriptsize{R$^*$}} (m-2-4);
    \path [>=latex,->] (m-2-4) edge (m-3-4);
    \path [>=latex,->] (m-4-3) edge (m-4-2);
     \draw [>=latex,->] (m-3-4) |- node[auto] {\scriptsize{ }} (m-4-3);
     \draw [>=latex,->] (m-1-4) -| (m-2-1);
     \draw [>=latex,->] (m-4-2) -| (m-2-1);
     \draw [>=latex,->] (m-2-1) |- (m-4-2);
     \draw [>=latex,->] (m-2-1) |- (m-1-4);
     \draw [>=latex,->] (m-2-4) edge (m-1-4);
\end{tikzpicture}

\caption{G-protein cycle, modified from \cite{Katanaev}. The activated receptor is denoted by R$^*$.}
\label{im:cycle}
\end{figure}

We consider the  (simplified) G-protein activation cycle shown in Figure~\ref{im:cycle}, see \cite{Katanaev,Leander, Woodroffe}. Once EP2 and EP4 are activated by PGE2 the G-protein complex G$_{\alpha\beta\gamma}$ consisting of an $\alpha$-, $\beta$- and $\gamma$-component can bind to it inside the cell.  The $\alpha$-component, in its inactive state,  is also bound to guanosine diphosphate (GDP). When the complex G$_{\alpha\beta\gamma}$ binds to the receptor the GDP-molecule dissociates and guanosine triphosphates (GTP) can bind to and activate the G-protein. This causes the G-protein complex to dissociate and to split into G$_{\beta\gamma}$ and the activated component G$_{\alpha}$. The activated G$_{\alpha}$ protein then hydrolyses GTP to GDP and becomes inactivated. Finally, G$_{\alpha}$ associates again with G$_{\beta\gamma}$ to form the complex G$_{\alpha\beta\gamma}$ and the cycle repeats.

There are four major families of G$_{\alpha}$-proteins, namely proteins with an $\alpha_s$, $\alpha_i$, $\alpha_q$ or $\alpha_{12/13}$-component, and different receptors activate different G$_{\alpha}$-proteins.  
EP2 only actives the G-protein G$_{\alpha_s}$ while EP4 activates both the G-proteins G$_{\alpha_s}$ and G$_{\alpha_i}$ \cite{Vleeshouwers}. 

To model the G-protein activation cycle we make the same simplifying assumptions as in \cite{Leander}, where the following reactions are labeled in  Figure~\ref{im:cycle},
\begin{align}
    1: &~~ \text{G}_{\alpha^{GDP}}+\text{G}_{\beta\gamma} \longleftrightarrow \text{G}_{\alpha^{GDP}\beta\gamma} \label{cycle eq1}\\
    2: &~~ \text{G}_{\alpha^{GDP}\beta\gamma} \longrightarrow \text{G}_{\alpha^{GTP}}+\text{G}_{\beta\gamma} \label{cycle eq2}\\
    3: &~~ \text{G}_{\alpha^{GTP}} \longrightarrow \text{G}_{\alpha^{GDP}} \label{cycle eq3}.
\end{align}
Namely, we assume that receptor activation and the association of receptors with G-proteins are rapid compared to G-protein activation, i.e.~the second reaction  takes place in a single step.
Moreover, we suppose that G$_{\alpha^{GTP}}$ has a negligible affinity for G$_{\beta\gamma}$ compared to the affinity of G$_{\alpha^{GDP}}$, i.e.~we assume that G$_{\alpha^{GTP}}$ does not bind to G$_{\beta\gamma}$.

\subsection{cAMP production}

The activity of EP2 and EP4 can be measured experimentally by determining cAMP levels. In the EP2 and EP4 signaling pathways
the cAMP concentration is mainly regulated by two enzymes, adenylyl cyclase (AC) and phospodiesterase (PDE) \cite{cAMP}. 
Depending on the cell type and receptors, different AC and PDE isoforms are expressed and specifically activated. AC and PDE isoforms show different regulation patterns \cite{ACisoforms},
but little is known about which specific isoforms are activated in the cAMP signaling cascade of EP2 and EP4. Table~\ref{ACisoform} shows the stimulation and inhibition of certain G-proteins for different AC isoforms. 

\begin{table}
    \centering
\begin{tabular}{ |p{3cm}||p{3cm}|p{3cm}|}
 \hline
 &\multicolumn{2}{|c|}{G-proteins} \\
 \hline
 AC-isoforms& Stimulation &Inhibition\\
 \hline
 Group I &&\\
    AC1& G$_{\alpha_s}$&G$_{\beta\gamma}$,G$_{\alpha_i}$,G$_{\alpha_z}$,G$_{\alpha_o}$\\
    AC8& G$_{\alpha_s}$&G$_{\beta\gamma}$\\
    AC3& G$_{\alpha_s}$&G$_{\beta\gamma}$\\\hline
    Group II&&\\
    AC2& G$_{\alpha_s}$, G$_{\beta\gamma}$&\\
    AC4& G$_{\alpha_s}$,G$_{\beta\gamma}$&\\
    AC7& G$_{\alpha_s}$,G$_{\beta\gamma}$&\\\hline
    Group III&&\\
    AC5&G$_{\alpha_s}$,G$_{\beta\gamma}$&G$_{\alpha_i}$,G$_{\alpha_z}$\\
    AC6&G$_{\alpha_s}$,G$_{\beta\gamma}$&G$_{\alpha_i}$,G$_{\alpha_z}$\\\hline
    Group IV&&\\
    AC9&G$_{\alpha_s}$&\\
 \hline
\end{tabular}
\caption{Regulation patterns of AC-isoforms \cite{ACisoforms}}
\label{ACisoform}
\end{table}

\begin{figure}[ht]
  \centering
    \includegraphics[height=100pt]{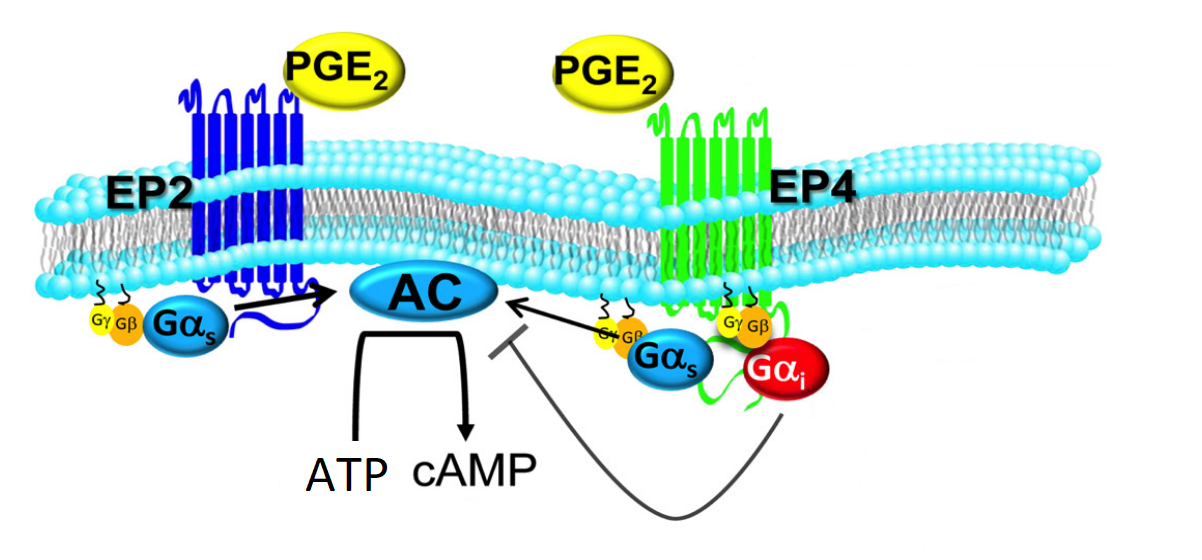}
    \caption{Schematic representation of the PGE2 signalling receptors EP2 and EP4, reproduced and modified from \cite{Keijzer}. EP2 and EP4 are GPCRs coexpressed in immune cells and specific for PGE2.  After binding to PGE2, they activate the G$_{\alpha s}$ proteins that in turn activate AC leading to increased levels of intracellular cAMP. EP4 can additionally activate G$_{\alpha i}$, leading to a modulation of cAMP production.}
    \label{im:cAMP}
\end{figure}

AC enzymes are activated by binding to an activated G$_{\alpha_s}$-unit. Once activated, AC converts adenosine triphosphate (ATP) into cAMP \cite{Hanoune}.  On the other hand, AC activity can be inhibited by active G$_{\alpha_i}$-units \cite{cAMP} as G$_{\alpha_i}$ bound to AC prevents binding to  G$_{\alpha_s}$ \cite{AC5}. 
EP2 activates G$_{\alpha_s}$ 
while
EP4 actives both, G$_{\alpha_s}$ and G$_{\alpha_i}$. Hence, EP4 may also have an inhibitory effect on cAMP  production, as shown in Figure \ref{im:cAMP}. 
The other important regulators of cAMP in the EP2 and EP4 signaling pathways are 
PDE enzymes. They catalyze the synthesis and decomposition of cAMP into adenosine monophosphate (AMP) and hence, lead to a decrease in cAMP levels \cite{cAMP}.

\subsection{Experimental data}

\begin{figure}[ht]
\includegraphics[width=.25\textwidth, height=3.7cm]{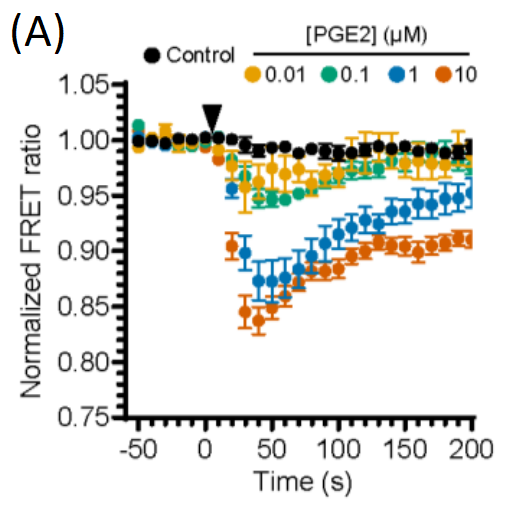}
\includegraphics[width=.75\linewidth, height=3.7cm]{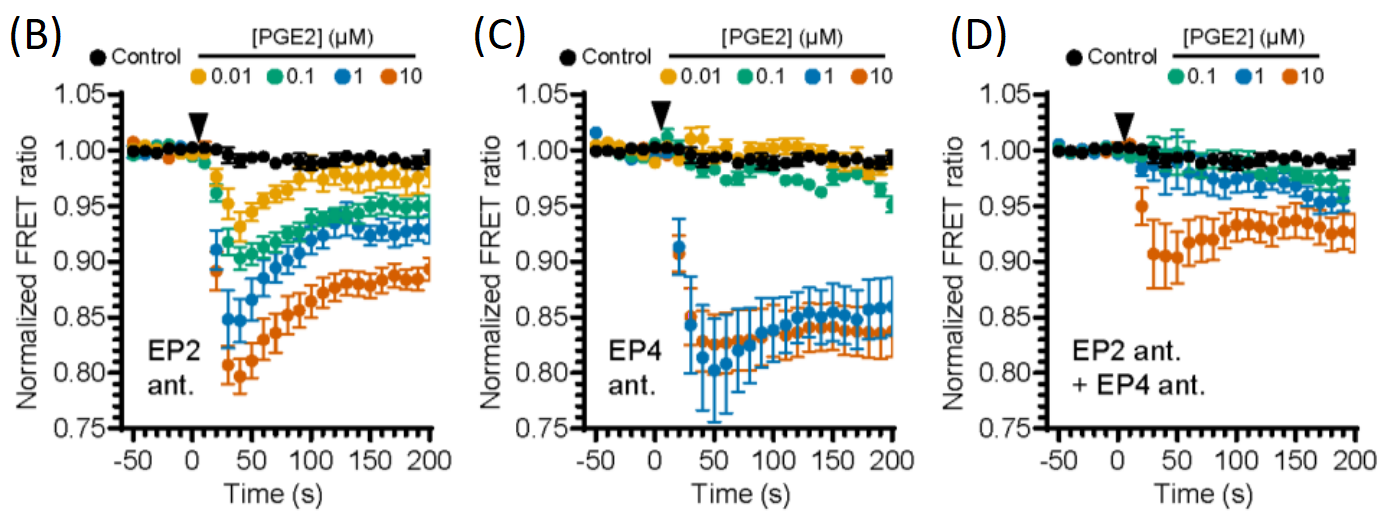}
\caption{Experimental data, reproduced from \cite{Vleeshouwers}.
EP2 and EP4 induce distinct cAMP responses. 
The intramolecular cAMP FRET sensor t-Epac-vv was used: binding of cAMP to t-Epac-vv reduces F\"orster Resonance Energy Transfer (FRET) between the mTurquoise donor and Venus acceptor fluorophores of t-Epac-vv, making a decreased ratio of the fluorescent intensities a direct measure of cAMP accumulation. (A) FRET ratios of t-Epac-vv before and after the addition of different PGE2 concentrations were measured in transiently transfected RAW macrophages. A control was performed with the addition of buffer only. The data presented are mean $\pm$ SD from $\geq 5$ cells
per condition.  FRET ratios were measured after the addition of PGE2 in cells pretreated with EP2 antagonist (ant.) AH6809 (B),  EP4 antagonist GW627368X (C), or pretreated with both GW627368X and AH6809 (D). The data presented are mean $\pm$ SD from $\geq 4$
cells per condition.}
\label{experiment}
\end{figure}

Figure \ref{experiment} shows data on EP2 and EP4 signaling in response to PGE2 based on experiments with mouse RAW 246.7 macrophages \cite{Vleeshouwers}. 
Here, the normalized FRET ratio is plotted against the time in seconds. The FRET ratio measures the transmission of energy from a donor molecule to an acceptor molecule. As the molecule binds to cAMP the FRET ratio decreases which indicates an increase in cAMP concentration. 
The FRET ratio was measured for the PGE2 concentrations $0.01\mu$M, $0.1\mu$M, $1\mu$M and $10\mu$M, as well as for a control experiment where only the buffer was added. 
The FRET ratio is shown for the following cases: 
\begin{align*}
&\text{(A) EP2 and EP4 active} \\
&\text{(B) EP4  active, EP2 blocked by an antagonist} 
\\
&\text{(C) EP2  active, EP4 blocked by an antagonist} \\
&\text{(D) EP2 and EP4 blocked by antagonists}
\end{align*}

It is important to remark that only relative cAMP concentrations are shown, and no quantitative measurements are available. The control experiment (black dots) is the baseline and the change in cAMP concentration is determined relative to that baseline.
In Figure \ref{experiment}(D) we observe cAMP production for a PGE2 concentration of 10~$\mu$M. A possible explanation is that the antagonists concentrations are too low for this high PGE level. This should be taken into account in Figures \ref{experiment}(A)--(C), i.e.~for PGE2 concentrations of~10$\mu$M, EP2 and/or EP4 may not be fully blocked by the antagonists.

\begin{figure}[ht]
\begin{center}
\includegraphics[height=7cm]{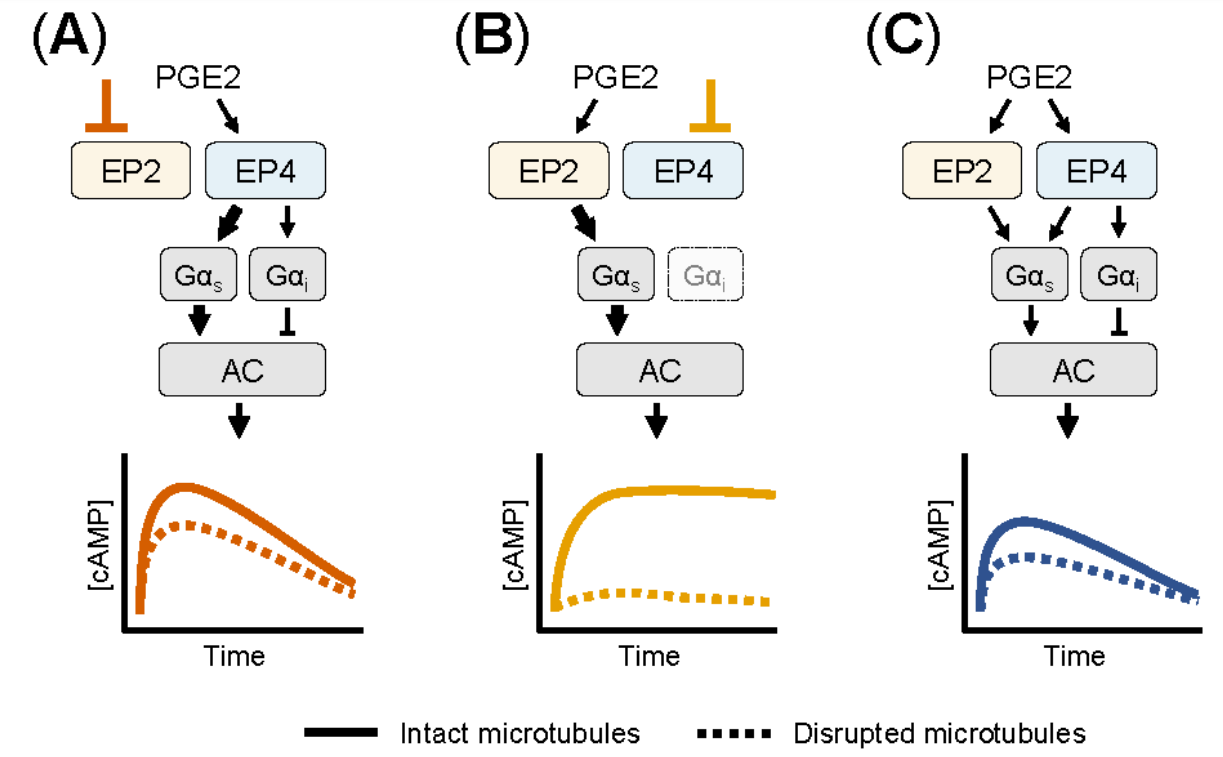}
\caption{Schematic overview of the cAMP responses induced by EP2 and EP4, reproduced from \cite{Vleeshouwers}. (A) When only EP4 is active, both G$_{\alpha s}$ and G$_{\alpha i}$ control AC activity. G$_{\alpha s}$ induces a dose-dependent cAMP response that is dampened by G$_{\alpha i}$. The cAMP signal subsides over time and is attenuated by microtubule disruption. (B) When EP2 is activated selectively, only G$_{\alpha s}$ modulates AC activity. The resulting cAMP response is either weak or strong, does not subside and completely relies on an intact microtubule network. (C) When both EP2 and EP4 are active, competition for G$_{\alpha s}$ dampens the integrated cAMP response. Signaling crosstalk between EP2 and EP4 allows the cell to respond differently to PGE2 depending on the organization and expression
of EP2 and EP4. }
\label{im:schematic}
\end{center}
\end{figure}

The data indicates significant qualitative differences in the signaling behavior of the receptors. Figure \ref{experiment}(C) displays that EP2 activity shows a clear threshold behavior with respect to PGE2. Namely, for the PGE2 concentrations 0.01$\mu$M and 0.1$\mu$M, there is negligible cAMP production, while for the PGE2 concentrations 1$\mu$M and 10$\mu$M, we observe a high production which stabilizes quickly. On the contrary, Figure \ref{experiment}(B) shows that cAMP production for EP4 increases gradually with PGE2 levels. Moreover, we notice a clear peak in the cAMP concentration 
 and then a decrease over time. The cAMP profiles for the combined signaling of EP2 and EP4 shown in Figure \ref{experiment}(A) are similar to EP4 but the cAMP levels are lower. Schematic plots are displayed in Figure~\ref{im:schematic}.

\section{Mathematical models}\label{sec:signEP2EP4}

In this section we develop mathematical models for the cAMP signaling cascades of EP2 and EP4 and their crosstalk. We address ligand-receptor binding, G-protein activation cycles and cAMP production separately.

\subsection{Ligand binding}\label{sec:lig}

EP2 and EP4 both bind to the ligand PGE2, but the receptors have distinct structural characteristics and show different signaling modes. 
To take these differences into account and to develop a model for the crosstalk of the receptors we describe the temporal dynamic interactions between ligands and receptors. 
We model the reactions in \eqref{rec_act} and denote the concentrations of EP2 by $r_1$, of EP4 by $r_2$ and the concentration of PGE2 by $p$. 
We assume that EP2 binds to PGE2 at rate $\mu_{1}$ and EP4 at rate $\mu_{2}.$ Once bound to PGE2 the receptors are in an activated state, denoted by $r_1^\ast$ and $r_2^\ast$, respectively. The dissociation rate for EP2 and PGE2 is $\delta_1$ and for EP4 and PGE2 $\delta_2$. 
EP4 shows internalization at rate $\rho_2$, while EP2 does not internalize. 
The dynamics of EP2 and EP4 activation through binding to the ligand PGE2 is then described by the system of ordinary differential equations 
\begin{align}\label{eq:lig_dyn}
\begin{split}
\frac{d p} {dt} & = - \mu_{m} \, r_m \,   p^{n_m}  + \delta_m r^\ast_m,  \\
\frac{dr_m}{dt} & =   - \mu_{m} \, r_m  p^{n_m}  + \delta_m r^\ast_m,   \\
\frac{d r^\ast_m}{dt} & = \mu_{m} \, r_m p^{n_m} - \delta_m r_m^\ast -\rho_mr_m^*,
\end{split}
\end{align}
where $m=1$ for EP2 and $m=2$ for EP4. The Hill coefficient for EP2 is $n_1\geq 3$ and for EP4 $n_2=1$.

The model reflects the qualitative differences between EP2 and EP4 in the ligand-receptor binding kinetics.
While EP4  undergoes rapid internalization
in response to PGE2, EP2 does not internalize~\cite{Desai, Vleeshouwers}, i.e.~$\rho_1=0, \rho_2>0$. Moreover, PGE2 has a higher affinity for EP4 than for EP2 which is reflected in the parameter values for the kinetic association and dissociation rates $\mu_m$ and $\delta_m$, $m=1,2$. 
Finally, EP2 shows ultrasensitivity with respect to PGE2 levels, i.e.\ a small change in ligand concentration can cause a large change in the concentration of activated G-proteins \cite{Roth}. This is modeled by a Hill coefficient $n_1\geq3$ and leads to the threshold behavior in cAMP production in response to PGE2 shown in Figure~\ref{experiment}. In \cite{Roth}, a Hill coefficient of at least $4$ is indicated to model ultrasensitivity. 

Different from related models for GPCR signaling pathways that assume ligand binding is rapid and use quasi-steady state assumptions  \cite{Leander,Woodroffe,Carvalho}, we specifically model ligand dynamics. 
System \eqref{eq:lig_dyn} is similar to the model for ligand binding to normal and decoy receptors in \cite{DeGhBa}, but adjusted to the characteristics of EP2 and EP4. More specifically, we take ultrasensitivity of EP2 into account by a Hill coefficient $n_1\geq3$ and the fact that EP4  undergoes rapid internalization
in response to PGE2 while EP2 does not internalize. 

Since EP2 does not internalize, the total EP2 receptor concentration  $r_1^{tot}=r_1+r_1^*$ is preserved. 
Assuming that ligand binding dynamics is rapid leads to the approximation
\begin{align}\label{eq:rapidbinding}
r^*_1=\frac{r_1^{tot}p^{n_1}}{K_D+p^{n_1}}, \quad \text{ where } \, K_D = \frac{\delta_1}{\mu_1} .
\end{align}
This quasi-steady state assumption with $n_1=1$ was used to model the EP2 cAMP signaling pathway in \cite{Leander}. However, a Hill coefficient $n_1=1$ reflects a gradual increase in cAMP levels and cannot predict the characteristic cAMP profiles in Figure \ref{experiment} displaying a threshold behavior with respect to PGE2 concentrations.
For EP2 the quasi-steady state assumption \eqref{eq:rapidbinding} with $n_1\geq 3$ leads to nearly the same cAMP profile for EP2 as the kinetic model \eqref{eq:lig_dyn}, but this assumption is not suitable for EP4.
In fact, taking ligand binding dynamics into account allows us to model internalization of EP4 which leads to a decay of cAMP expression over time, as well as to combine the models for the EP2 and EP4 signaling pathways and describe  crosstalk of the receptors.

\subsection{EP2 G-protein activation cycle}\label{sec:EP2}

EP2 only activates the G-protein G$_{\alpha_s}$. 
We consider the G-protein activation cycle in Figure~\ref{im:cycle} and 
denote by $V_1$ the reassociation and dissociation rate of G$_{\alpha_s\beta\gamma}$ in  (\ref{cycle eq1}), by $V_2$  the dissociation rate of G$_{\alpha_s\beta\gamma}$ catalyzed by the activated receptor EP2$^*$ in  (\ref{cycle eq2}) and by $V_3$ the hydrolyses rate of GTP on G$_{\alpha_s}$ in (\ref{cycle eq3}). Denoting the concentrations of G$_{\alpha_s\beta\gamma}$, G$_{\beta\gamma}$, G$_{\alpha_s^{GTP}}$ and G$_{\alpha_s^{GDP}}$ by $\beta_{s}$, $\beta$, $\alpha_s^{*}$ and $\alpha_s$, respectively, the G-protein activation cycle for EP2 is modeled by the system of ordinary differential equations
\begin{align*}
    \frac{d\beta_{s}}{dt} &= V_1 - V_2,&
     \frac{d\alpha_s^*}{dt} &= V_2 - V_3 ,\\
    \frac{d\beta}{dt} &= V_2 - V_1 ,&
    \frac{d\alpha_s}{dt} &= V_3 - V_1 .
\end{align*}
Based on the Law of Mass Action for the rate $V_1$ we have 
\begin{align*}
    V_1= k_{\beta s}\alpha_s\beta - d_{\beta s}\alpha_s\beta,
\end{align*}
where $k_{\beta s}$ denotes the G$_{\alpha_s}$-$\beta\gamma$ association rate and $d_{\beta s}$ the G$_{\alpha_s\beta\gamma}$ dissociation rate. The protein complex 
G$_{\alpha_s\beta\gamma}$ binds to the activated receptor and then dissociates into G$_{\alpha_s^{GTP}}$ and G$_{\beta\gamma}$. We assume that the binding of GTP is rapid compared to the dissociation of GDP, so that the receptor catalyzes the transition from G$_{\alpha_s^{GDP}\beta\gamma}$ to G$_{\alpha_s^{GTP}}$ in a single step.  
This enzymatic reaction is modelled by the Michaelis-Menten rate equation 
\begin{align*}
    V_2 = k_s r_1^*\frac{\beta_{s}}{\beta_{s}+ K_{r_1s}},
\end{align*}
where $k_s$ is the G$_{\alpha_s}$ activation rate, $r_1^*$ the concentration of activated EP2 receptors determined by \eqref{eq:lig_dyn} and  $K_{r_1s}$ the dissociation constant for EP2 and G$_{\alpha_s\beta\gamma}$.
Finally, the hydrolysis of GTP to GDP is described by
\begin{align*}
    V_3 = d_s\alpha_s^*,
\end{align*}
where $d_s$ is the hydrolysis rate.
Adding the respective equations we observe that the following relations for mass conservation hold,
$$
    \beta+\beta_s=b_1,\qquad 
    \alpha_s + \beta_{s} + \alpha_s^* = a_s,
$$
where $b_1$ is the total concentration of G$_{\beta\gamma}$-units and $a_s$ the total concentration of G$_{\alpha_s}$-units.
Using these relations leads to the reduced system
\begin{align}\label{EP2:model1}
\begin{split}
 \frac{d\alpha_s^\ast}{dt} &= k_{s}r^\ast_1 \frac{ b_1 - \beta}{b_1 - \beta + K_{r_1s}} - d_s \alpha^\ast_s,\\
 \frac{d\beta}{dt}& =k_{s}  r^\ast_1 \frac{ b_1 - \beta}{b_1 - \beta + K_{r_1s}} + d_{\beta s}(b_1 - \beta) - k_{\beta s} \beta(a_s - b_1 + \beta - \alpha^\ast_s).
 \end{split}
\end{align}
System \eqref{EP2:model1} is based on the  model for the G-protein activation cycle of EP2 in \cite{Leander}. However, here, we take ligand binding kinetics and ultrasensitivity into account, i.e.\ 
$r^*_1$ denotes the concentration of activated EP2 receptors modeled by system~\eqref{eq:lig_dyn}. 
In \cite{Leander} ligand binding was assumed to be rapid and the quasi steady state approximation  \eqref{eq:rapidbinding} with Hill coefficient $n_1=1$ was used. However, as discussed in the previous subsection taking $n_1=1$ does not resemble the threshold behavior of EP2 in response to PGE2, see Figure~\ref{experiment}. Moreover, the ligand-receptor dynamics is essential for EP4 and for modeling crosstalk of the receptors. 

Several models for the G-protein activation cycle of general GPCRs have been developed, with varying level of detail and complexity. For an overview we refer to \cite{Carvalho,KiLi, Woodroffe,Leander}. Most models describe general mechanisms, but specific models for the G-protein activation cycles for EP2 and EP4 are not available. An exception is \cite{Leander} where the EP2 cAMP signaling pathway was modeled. Finally we remark that as in \cite{Leander} we do not 
assume that the total concentration of G$_{\beta\gamma}$-units $b_1$ and G$_{\alpha_s}$-units $a_s$ coincide, which differs from most models \cite{Bridge, Katanaev,Woodroffe}. Only few studies have quantified the ratio between the total G-protein concentrations $b_1$ and  $a_s$, but evidence suggests that it is not equal to one \cite{Leander}.

\subsection{EP4 G-protein activation cycles}\label{sec:EP4}

EP4 activates two different G-proteins,  G$_{\alpha_s}$ and G$_{\alpha_i}$. We assume that EP4 responds to both G-proteins in the same way and activation of G$_{\alpha_s}$ and G$_{\alpha_i}$ occurs according to the G-protein  activation cycle shown in Figure~\ref{im:cycle}. 
Furthermore, we assume that the protein complexes G$_{\alpha_s\beta\gamma}$ and G$_{\alpha_i\beta\gamma}$ are competing for EP4.
We use the same notations as in Section \ref{sec:EP2} for the rates $V_1,V_2$ and $V_3$ describing the G$_{\alpha_s}$ activation cycle, replacing EP2 by EP4. 
The association and dissociation rate of G$_{\alpha_i\beta\gamma}$ is denoted by $V_4$, the dissociation rate of G$_{\alpha_i\beta\gamma}$ catalyzed by the activated receptor EP4$^*$ by $V_5$ and the hydrolyses rate of GTP on G$_{\alpha_i}$ by $V_6$. The G$_{\alpha_s}$ and G$_{\alpha_i}$ activation cycles for EP4 are then described by the system of ordinary differential equations
\begin{equation}\label{eq:ratesEP4}
\begin{aligned}
    \frac{d\beta_{s}}{dt} &= V_1 - V_2, &  \frac{d\beta_{i}}{dt} &= V_4 - V_5,\\
    \frac{d\alpha_s^*}{dt} &= V_2 - V_3, & 
    \frac{d\alpha_i^*}{dt} &= V_5 - V_6, \\    
    \frac{d\alpha_s}{dt} &= V_3 - V_1, &
    \frac{d\alpha_i}{dt} &= V_6 - V_4,\\
    \frac{d\beta}{dt} &= V_2 - V_1 + V_5 - V_4,
\end{aligned}
\end{equation}
where $\beta_{i}$, $\alpha_i^{*}$ and $\alpha_i$ are the concentrations of G$_{\alpha_i\beta\gamma}$,  G$_{\alpha_i^{GTP}}$ and G$_{\alpha_i^{GDP}}$, respectively.
The rates $V_1$ and $V_3$ for the G$_{\alpha_s}$-activation cycle  are the same as for EP2, and correspondingly, for G$_{\alpha_i}$ we have
$$
V_4= k_{\beta i}\alpha_i\,\beta - d_{\beta i}\beta_{i},\qquad V_6 = d_i\alpha_i^*,
$$
where $k_{\beta i}$ denotes the G$_{\alpha_i}$-$\beta\gamma$ association rate, $d_{\beta i}$ the G$_{\alpha_i\beta\gamma}$ dissociation rate and $d_i$ the hydrolysis rate.
In the G-protein activation by EP4 we take competitive binding into account, i.e.~we assume that if the complex G$_{\alpha_s\beta\gamma}$ is bound to the receptor it cannot bind to G$_{\alpha_i\beta\gamma}$, and vice versa.
This is modelled by the rates
\begin{align*}
    &V_2 = k_sr_2^*\frac{\beta_{s}}{\beta_{s}+ K_{r_2s}+ \tfrac{K_{r_2s}}{K_{r_2i}}\beta_s},
    \qquad V_5= k_ir_2^*\frac{\beta_{i}}{\beta_{i}+ K_{r_2i} + \tfrac{K_{r_2s}}{K_{r_2i}}\beta_i},
\end{align*}
where $k_s$ and $k_i$ are the G$_{\alpha_s}$ and  G$_{\alpha_i}$ activation rates, $r_2^*$ the concentration of activated EP4 receptors determined by \eqref{eq:lig_dyn}  and $K_{r_2s}$ and 
$K_{r_2i}$ the EP4 dissociation constants for  G$_{\alpha_s\beta\gamma}$ and 
G$_{\alpha_i\beta\gamma}$, respectively. These rates are an approximation for competitive binding which is derived as follows. For the concentration of activated EP4 receptors $r_2^*$ we have 
\begin{align}\label{eq:approximation}
r_2^*=r_{2s}^*+r_{2i}^*+r_{20}^*,    
\end{align}
where $r_{2s}^*$ and $r_{2i}^*$ are the concentrations of activated EP4 bound to G$_{\alpha_s\beta\gamma}$ and 
G$_{\alpha_i\beta\gamma}$, respectively, and $r_{20}^*$ is the concentration of activated EP4 unbound to G-protein complexes. We model  the binding of the G-protein complexes to the receptor by the Michaelis-Menten rate equations
$$
r_{2s}^*=\frac{r_{20}^*\beta_s}{K_{r_2s}+r_{20}^*}, \qquad 
r_{2i}^*=\frac{r_{20}^*\beta_i}{K_{r_2i}+r_{20}^*}.
$$
Assuming that $K_{r_2s} \gg r_{20}^*$ and $K_{r_2i}\gg r_{20}^*$ we can approximate these rates by
$$
r_{2s}^*\approx\frac{r_{20}^*\beta_s}{K_{r_2s}}, \qquad 
r_{2s}^*\approx\frac{r_{20}^*\beta_i}{K_{r_2i}}.
$$
Using these relations in \eqref{eq:approximation} we obtain
$$
r_{20}^*\approx\frac{r_2^*}{1+\frac{\beta_s}{K_{r_2s}}+\frac{\beta_i}{K_{r_2i}}},
$$
which leads to the approximate rates used in $V_2$ and $V_5$, 
$$
r_{2s}^*\approx
\frac{r_2^*\beta_s}{\beta_s+K_{r_2s}+\frac{K_{r_2s}}{K_{r_2i}}\beta_s}, \qquad 
r_{2i}^*\approx \frac{r_2^*\beta_{i}}{\beta_{i}+ K_{r_2i} + \tfrac{K_{r_2s}}{K_{r_2i}}\beta_i}.
$$

Adding the corresponding equations in System \eqref{eq:ratesEP4}
we observe that the following mass conservation assumptions hold
$$
    \beta+\beta_s+\beta_i=b_2,\qquad 
    \alpha_s+\beta_s+\alpha_s^*=a_s,\qquad 
    \alpha_i+\beta_i+\alpha_i^*=a_i,
$$
where $b_2$ is the total concentration of G$_{\beta\gamma}$-units and  $a_s$ and $a_i$ are the total concentrations of G$_{\alpha_s}$-units and G$_{\alpha_i}$-units respectively.
Using these relations we can reduce the model to the system 
\begin{align}\label{EP4:model3}
\begin{split}
\frac{d\alpha_s^\ast}{dt} &= k_{s} r^*_2 
\frac{ b_2 - \beta- \beta_i}{b_2 - \beta  - \beta_i+ K_{r_2s} + \tfrac{K_{r_2s}}{K_{r_2i}} \beta_i} - d_s \,  \alpha_s^\ast,\\
\frac{d\alpha_i^\ast}{dt}&= k_{i} r^*_2 \frac{\beta_i}{\beta_i + K_{r_2i} + (b_2 - \beta - \beta_i)\tfrac{K_{r_2i}}{K_{r_2s}}} - d_i \,  \alpha_i^\ast,\\
\frac{d\beta_i}{dt}& =- k_{i} r^\ast_2 \frac{\beta_i}{\beta_i + K_{r_2i} + (b_2 - \beta - \beta _i)\tfrac{K_{r_2i}}{K_{r_2s}}}  + k_{\beta i} \beta ( a_i - \beta_i - \alpha_i^\ast) - d_{\beta i} \, \beta_i, \\
\frac{d\beta}{dt} & = k_{i} r^\ast_2 \frac{\beta_i}{\beta_i + K_{r_2i} + (b_2 - \beta - \beta_i)\tfrac{K_{r_2i}}{K_{r_2s}}}
 + k_{s} r^*_2 
\frac{ b_2 - \beta- \beta_i}{b_2 - \beta  - \beta_i+ K_{r_2s} + \tfrac{K_{r_2s}}{K_{r_2i}} \beta_i}\\ 
 + &d_{\beta i} \beta_i +  d_{\beta s} (b_2 - \beta- \beta_i) - \beta(k_{\beta s}(a_s - b_2 + \beta + \beta_i -\alpha_s^\ast) + k_{\beta i}(a_i - \beta_i - \alpha_i^\ast)).
\end{split}
\end{align}

Most models for G-protein activation cycles of GPCRs describe the binding and activation of one specific G-protein \cite{Bridge, Katanaev, Leander,Woodroffe}. While models for the binding of two different ligands to a receptor have been considered, such as agonist–antagonist competition \cite{BridgeKing},  we are not aware of models for GPCRs that take the activation of two different G-proteins into account. 
System \eqref{EP4:model3} describes EP4, a GPCR activating two G-proteins and extends the single G-protein activation model for EP2 \eqref{EP2:model1}. It models the EP4 activation cycles for G$_{\alpha_s}$ and G$_{\alpha_i}$  and takes the effects of competitive binding between G$_{\alpha_s}$ and G$_{\alpha_i}$ into account.

\subsection{cAMP production}\label{sec:cAMP}

The production of cAMP in the EP2 and EP4 signaling pathways is mainly regulated by two enzymes, AC and PDE.
There exist different isoforms of AC and PDE and depending on the cell type and organism several of them are present in a cell. For the different regulation patterns of AC-isoforms we refer to Table~\ref{ACisoform}. 
AC activation and cAMP degradation by PDE in response to EP2 and EP4 is not yet very well understood. For instance, it is known that both receptors activate AC2, but not AC3 \cite{Bogard}.
In cells similar to mouse RAW 246.7 macrophages the isoforms AC2, AC3, AC6, AC7, and AC9 were found \cite{Jiang, Tarnawski}. Concerning the degradation by PDEs, the isoform PDE4 is considered the principle enzyme constraining cAMP signaling \cite{Bogard,Leander}. 

Since information about the specific AC and PDE isoforms is not available and to reduce the number of unknown parameters in the model, we assume that one dominant AC-isoform is responsible for cAMP production, for which G$_{\alpha_s}$ has a stimulating and G$_{\alpha_i}$ an inhibiting effect, and that the isoform PDE4 dominates cAMP degradation.

For EP2 that only activates~G$_{\alpha_s}$
we use a Michaelis-Menten reaction function to model AC activation by activated~G$_{\alpha_s}$, 
\begin{align*}
     \frac{\alpha_s^*A^{tot}}{\alpha_s^*+K_{as}},
\end{align*}
where $A^{tot}$ denotes the total AC concentration and
$K_{as}$ the dissociation constant for AC and G$_{\alpha_s}$. 
EP4 does not only activate  G$_{\alpha_s}$, but also  G$_{\alpha_i}$ which has an inhibiting effect on cAMP production. More specifically,
G$_{\alpha_i}$ proteins can bind to AC preventing the binding of G$_{\alpha_s}$ to AC and hence, inhibit cAMP production. To take inhibition by G$_{\alpha_i}$ into account we use the same approximation as in Section 
\ref{sec:EP4} for the competitive binding of 
G$_{\alpha_s\beta\gamma}$ and  G$_{\alpha_i\beta\gamma}$ to EP4.
Denoting by $AC$ the concentration of AC unbounded to activated G-proteins we have 
$$
AC^{\rm tot} = \frac{AC \alpha_s^\ast}{K_{as} + AC} + AC + \frac{ AC\alpha_i^\ast }{K_{ai} + AC},
$$
where $K_{ai}$ is the dissociation constant for AC and G$_{\alpha_i}$.
Assuming that $K_{as} \gg AC$ and $K_{ai} \gg AC$  leads to the approximation
$$
AC^{\rm tot} \approx \frac{AC \alpha_s^\ast}{K_{as} } + AC + \frac{ AC\alpha_i^\ast }{K_{ai} },
$$
and we obtain
$$
AC= \frac{AC^{\rm tot}}{ 1+ \alpha_s^\ast/K_{as} + \alpha_i^\ast/K_{ai}}.
$$
Hence, as in Section~\ref{sec:EP4},
the  approximate cAMP production rate is given by
$$
\frac{ \alpha^\ast_sA^{tot}}{\alpha^\ast_s + K_{as} (1+ \alpha^\ast_i /K_{ai})},
$$
which takes inhibition by G$_{\alpha_i}$ into account.
PDE enzymes catalyze the decomposition of cAMP into AMP which we model by 
$$
   \frac{wP^{tot}}{w+K_w},
$$
where $w$ denotes the cAMP concentration,  $P^{tot}$ is the total PDE4 concentration  and $K_w$ is the dissociation constant for PDE4 and cAMP.   
Since relative concentrations are measured in the experiments, see Figure~\ref{experiment}, we assume that there is no basal cAMP production rate. 
Hence, cAMP production is described by the ordinary differential equation
\begin{align}\label{eq:EP2cAMP}
 \frac{dw}{dt} & = k_{w}\frac{ \alpha^\ast_sA^{tot} }{\alpha^\ast_s+K_{as}+\eta K_{as}\tfrac{\alpha^\ast_i}{K_{ai}}}  - d_{w} P^{tot}\frac{ w}{w+K_{w}}, 
\end{align}
where $k_{w}$ is the active cAMP production rate and $d_w$ the PDE4 catalyzed cAMP degradation rate. Moreover, through the parameter $\eta$ we can take inhibition by G$_{\alpha_i}$ into account, i.e.\ we set $\eta=0$ for EP2 and $\eta=1$ for EP4.

There exist several models for cAMP regulation in GPCR signaling pathways, most are more detailed and take additional processes into account \cite{Leander,Saucerman,Ohadi,Xin}. Here, we only consider the main enzymes,  AC and PDE, responsible for cAMP expression in EP2 and EP4 signaling. We make the simplifying assumption that for each of these enzymes there is one dominant isoform.
We only have qualitative  experimental data and no information is available about the specific isoforms  that are responsible for the cAMP signaling pathways of EP2 and EP4 which would be required for a more refined modeling approach. Different from related models for cAMP expression such as \cite{Leander,Saucerman,Xin} no basal production rate is included in  \eqref{eq:EP2cAMP} since the experimental data display cAMP profiles relative to a control experiment where no ligand is added. 
We took the inhibiting effect of G$_{\alpha_i}$ into account based on competitive binding which is modeled in the same way as the competitive binding of G$_{\alpha_s}$ and G$_{\alpha_i}$ to EP4. 
Instead of an approximation as we use here an explicit expression for the effects of competitive binding is used in  \cite{Wang}.
Another more detailed model for inhibition by G$_{\alpha_i}$ including the possibility of dimerization and of the simultaneous binding of two G$_\alpha$ molecules to AC5 was proposed in \cite{Goodspeed}.

\subsection{Parameter values}

For only some of the parameters in equations~\eqref{eq:lig_dyn}, \eqref{EP2:model1}, \eqref{EP4:model3} and \eqref{eq:EP2cAMP} precise values are available for the receptors EP2 and EP4 and the specific cell types used in the experiments.  The values used in the simulations in Section~\ref{sec:sim} are listed in Tables~\ref{table:1}-\ref{table:3}. If a parameter range is given  the value used is  indicated in brackets. 

\subsubsection{Ligand receptor binding}

To model the threshold  behavior of EP2 with respect to PGE2 we assume that EP2 shows ultrasensitivity and take the Hill coefficient $n_1=3$. For EP4 cAMP levels increase gradually with respect to PGE2 concentrations and hence, for EP4 we take $n_2=1$.
While EP2 does not internalize, i.e. $\rho_1=0$, internalization plays a crucial role in the ligand binding dynamics for EP4  \cite{Desai, Hata} and leads to the decrease of cAMP levels over time in Figure \ref{experiment}. The internalization rate $\rho_2$ is calculated based on~\cite{Desai} where  internalization of EP4 was measured when exposed to $1\mu$M PGE2.
It was found that EP4 receptors underwent rapid internalization to the extent of 40\% with a half-time of
approximately $2$~minutes. 

EP4 typically shows a higher binding affinity to PGE2 than EP2. Based on \cite{MaJaSoMl} the equilibrium dissociation constants $K_D$  are $12$nM for PGE2 and EP2 and $1.9$nM for PGE2 and EP4 (mouse). 
Taking ligand binding dynamics into account requires the kinetic association and dissociation rates for ligand and receptor, but most studies focus on the equilibrium dissociation constants $K_D$ or inhibitor constants $K_i$ \cite{MaJaSoMl}. The  kinetic association and dissociation rates are not available in the literature for EP2 and EP4 and the cell type used in the experiments. We therefore use the association and dissociation rates
for PGE2 and EP4 in \cite{Kurz} for human embryonic kidney cells.
For PGE2 and EP2 kinetic association and dissociation rates are not available and therefore, we use the dissociation rate $\delta_1$ for DcR3 receptors and the ligand CD95L in \cite{DeGhBa} and calculate the association rate $\mu_1$ based on the relation 
$K_D=\delta_1/\mu_1$ and the value $K_D=12$nM for PGE2 and EP2 in \cite{MaJaSoMl}.
We remark that for EP4
the relation $K_D=\delta_2/\mu_2$ and parameter values in \cite{Kurz} lead to $K_D=0.02401\mu$M, which is an order of magnitude higher than the rate $K_D=1.9$nM in \cite{MaJaSoMl}. Effects due to internalization could explain this difference, since the rate $K_D$ reflects an equilibrium dissociation rate. 

The total receptor concentration $r_1^{tot}$ for EP2  is taken from \cite{LeFrcor, Katanaev}. Moreover, we assume that the total concentration of EP4 is equal or twice as high as of EP2, as indicated in~\cite{Vleeshouwers}.  
To model crosstalk  we combine the models for the ligand receptor binding for EP2 and EP4.  Both receptors bind to the same ligand PGE2. We assume that the affinity of 
PGE2 for EP2 is the same as for EP4 and hence, take as binding ratio for crosstalk $\sigma=0.5$.

\begin{table}[ht]
\centering
\begin{tabular}{ |c| c| c|c| }
\hline
symb.& parameter& value & reference\\
\hline
$\mu_1$ & assoc.~rate EP2 - PGE2 & $0.2417 (s\mu M)^{-1}$  & \cite{MaJaSoMl} (calculated from $K_d$, mouse) \\
$\mu_2$ & assoc.~rate EP4 - PGE2 & $1.385 (s\mu M)^{-1}$ & \cite{Kurz} (human embryo.~kidney cells) \\
$\delta_1$ & dissoc.~rate EP2 - PGE2 & $0.0029\, s^{-1}$ & \cite{DeGhBa} (CD95L, DcR3 decoy receptor)  \\
$\delta_2$ & dissoc.~rate EP4 - PGE2& $0.03325\, s^{-1}$ & \cite{Kurz} (human embryo.~kidney cells)   \\
$\rho_1$ & internalization rate EP2& $0$ &  assumed\\
$\rho_2$ & internalization rate EP4 & $0.00538$  & \cite{Desai} (estimat., human 293-EBNA)\\
$n_1/n_2$ & Hill coefficient EP2 /EP4 & $3$/$1$ & assumed \\
$r_1^{tot}$ &total $EP2$ concentration & $2$--$5$nM~(4nM)  &  \cite{LeFrcor, Katanaev} (human T-cells)\\
$r_2^{tot}$ &total $EP4$ concentration & $8$nM   &assumed\\
$\sigma$ & binding ratio for crosstalk & $0.5$  & assumed\\
\hline
\end{tabular}
\caption{Parameter values for ligand dynamics and receptor activation}
\label{table:1}
\end{table}

As initial values in the simulations we take the PGE2 concentrations used in the experiments and assume that initially, no receptor is activated, i.e.,
\begin{align*} 
p(0) \in\{ 0.01\mu{\rm M}, 0.1\mu{\rm M}, 1\mu M, 10\mu{\rm M}\}, \qquad r_m^\ast(0)=0, \quad r_m(0)=r_m^{tot}, \quad \text{ for } \;  m=1,2, 
 \end{align*}

\subsubsection{G-protein activation cycles}

The model for the G-protein activation cycle of EP2 \eqref{EP2:model1} and the extended model for EP4 \eqref{EP4:model3} are based on the model for the signaling pathways of EP2 and C5a in \cite{Leander}. We take the parameter values in this reference for EP2 and the reaction rates in the G-protein activation cycles. The dissociation constants for EP4 and the G-protein complexes 
G$_{\alpha_s\beta\gamma}$ and G$_{\alpha_i\beta\gamma}$ are taken from~\cite{KiKiShKiYu}. As in \cite{Leander} we assume that the association rates for $G_{\alpha_s}$ and $G_{\beta\gamma}$ and for $G_{\alpha_i}$ and $G_{\beta\gamma}$ are the same. 
For EP4 inhibition by G$_{\alpha_i}$ is taken into account in \eqref{EP4:model3} through competitive binding and determined by the ratio of the dissociation constants $K_{r_2s}/K_{r_2i}$.

We remark that the total G$_{\beta\gamma}$ concentration $\beta\gamma^{tot}$ in \cite{Leander} is based on the 
mass conservation assumption
\begin{equation*}
    \beta+\beta_s+\beta_i=\beta\gamma^{tot}.
\end{equation*}
As EP2 does not activate the G-protein G$_{\alpha_i}$ we neglect G$_{\alpha_i\beta\gamma}$ in model~\eqref{EP2:model1}
and therefore use a lower value for $\beta\gamma^{tot}$ for EP2. 
Here, we use $b_1=\beta\gamma^{tot}=5$nM as it matches the initial value for G$_{\alpha_i\beta\gamma}$ taken for EP4 while we take $b_2=\beta\gamma^{tot}=1.8\mu$M for EP4. 

To model crosstalk of the receptors we combine the models \eqref{EP2:model1} and \eqref{EP4:model3}. Both receptors activate the G-protein G$_{\alpha_s}$. We assume that the affinity of 
G$_{\alpha_s}$ for the receptors EP2 and EP4 is the same, leading to the binding ratio $\sigma_0=0.5$.

\begin{table}[h!]
\centering
\begin{tabular}{ |c| c| c|c| }
\hline
symb.& parameter& value & reference\\
\hline
$k_{s}$ & $G_{\alpha_s}$ activation rate& $1$--$5 s^{-1}\ (5 s^{-1})$ & \cite{Leander}  ($\beta$-adrenergic recept.) \\
$k_{i}$   & $G_{\alpha_i}$ activation rate& $5 s^{-1}$ &  \cite{Leander}  ($\alpha_{2a}$ adreno recept.)\\
$d_s$    & $G_{\alpha_s}$ hydrolysis rate& $0.04, 0.07 s^{-1} (0.07s^{-1})$ &\cite{Leander}  \\
$d_i$    & $G_{\alpha_i}$ hydrolysis rate& $0.03 s^{-1}$ & \cite{Leander}  \\
$k_{\beta s}$  & $G_{\alpha_s}$-$\beta\gamma$ assoc. rate& $0.7 {\rm \mu M}^{-1} s^{-1}$  &  \cite{Leander} (estimate)\\
$k_{\beta i}$   & $G_{\alpha_i}$-$\beta\gamma$ assoc.~rate  & $0.7{\rm \mu M}^{-1} s^{-1}$ &  \cite{Leander} (rat, bovine)\\
$d_{\beta s}$   &$G_{\alpha_s\beta\gamma}$ dissoc.~rate & $18.9 \cdot 10^{-3} s^{-1}$  &\cite{Leander} (estimate)\\
$d_{\beta i}$  &$G_{\alpha_i\beta\gamma}$ dissoc.~rate & $0.14 \cdot 10^{-3} s^{-1}$   & \cite{Leander} (estimate) \\
$K_{r_1s}$  & EP2-$G_{\alpha_s\beta\gamma}$ dissoc.~const.& $0.8\mu M$  & \cite{Leander} ($\beta$-adrenergic recept.) \\
$K_{r_2s}$  & EP4-$G_{\alpha_s\beta\gamma}$ dissoc.~const.& $0.088\mu M$  & \cite{KiKiShKiYu} (human)\\
$K_{r_2i}$  & EP4-$G_{\alpha_i\beta\gamma}$ dissoc.~const.& $50, 175, 214 nM$  ($214nM$)& \cite{KiKiShKiYu} (human)\\
$a_s$ & total $G_{\alpha_s}$ concentration &  $2.3 \mu M$& \cite{Leander} (myocyte)\\
$a_i$ & total $G_{\alpha_i}$ concentration& $8 \mu M$ & \cite{Leander} (neutrophils)\\
$b_1$ & total $G_{\beta\gamma}$ concentr.~EP2 & $0.005 \mu M$  & \cite{Leander} (neutrophils, estim.) \\
$b_2$ & total $G_{\beta\gamma}$ concentr.~EP4 & $1.8 \mu M$  & \cite{Leander} (estimate) \\
$\sigma_0$ & binding ratio for crosstalk & 0.5/0.7  & assumed\\
\hline
\end{tabular}
\caption{Parameter values for the G-protein activation cycles}
\label{table:2}
\end{table}

As initial values  we take the basal steady-state concentrations in absence of PGE2 obtained by running the corresponding model in the absence of PGE2,
$$
\alpha_s^*(0)=0,\qquad \beta(0)=0.06 nM,
$$
and for EP4 we have, in addition, 
$$
\alpha_i^*(0)=0, 
 \qquad \beta_i(0)=1.795 \mu M.
$$

\subsubsection{cAMP production}

The experimental data show inhibition by G$_{\alpha_i}$ in the cAMP signalling cascade of EP4~\cite{Vleeshouwers}. Among the AC-isoforms found in cells similar to mouse RAW 246.7 macrophages, AC$6$ is the only one that is inhibited by G$_{\alpha_i}$, see Table~\ref{ACisoform}. Hence, 
we assume that AC$6$ is the dominant isoform for cAMP production. The parameter values for AC$6$ are taken from~\cite{Goodspeed}. Here, we assume that AC$6$ has the same molecular weight as AC$2$ given in~\cite{Leander}, namely $106$~kDa. We also assume as in~\cite{Leander}, that AC$6$ constitutes $0.1$\%   of transfected S$49$ membranes which allows to rewrite the active cAMP production from $3.8$ nmol/min~mg into $6.713$~s$^{-1}$. 
Concerning cAMP degradation by PDEs we assume that PDE4 is the principle isoform constraining cAMP expression. The parameter values for PDE4 are taken from~\cite{Ohadi}.

\begin{table}[ht]
\centering
\begin{tabular}{ |c| c| c|c| }
\hline
symb.& parameter& value & reference\\
\hline
$k_w$ & active cAMP production rate & $6.713 s^{-1}$   & \cite{Leander} (modif.~for~AC6)\\
$K_{as}$    & dissociat.~const for AC6 and $G_{\alpha_s}$& $0.2\mu M$ & \cite{Goodspeed}\\
$K_{ai}$      & dissociat.~const for AC6 and $G_{\alpha_i}$ & $0.027\mu M$  & \cite{Goodspeed}\\
 $d_{w}$  & cAMP degradation rate by PDE4& $8.66 s^{-1}$ & \cite{Ohadi} \\
$K_{w}$     & dissociat.~const for PDE4 and cAMP& $1.21\mu M$  & \cite{Ohadi}\\
$A^{tot}$ & total AC concentration&   $0.0497\mu M$ &\cite{Saucerman}\\
$P^{tot}$ &total PDE concentration & $0.039\mu M$ &\cite{Saucerman} \\
\hline
\end{tabular}
\caption{Parameter values for cAMP production}
\label{table:3}
\end{table}


We assume that initially, no receptor is activated and hence no cAMP is in the system, since cAMP levels are measured relative to a control experiment, i.e.\ as initial value we take
$$
w(0)=0.
$$

\section{Simulations and results}\label{sec:sim}

In Subsection~\ref{sec:full} we present the full models for the cAMP signaling pathways for EP2, EP4 and their crosstalk. Subsequently,  we show in numerical simulations that the models qualitatively predict the experimentally observed cAMP levels in response to PGE2 shown in Figures~\ref{experiment} and~\ref{im:schematic}.  Numerical codes for simulations of model equations are implemented
in Python, taking advantage of the Scipy module \cite{Jones2001} and  solutions for systems of ODEs were obtained using the scipy.integrate.odeint package.

\subsection{Full models for signaling pathways}\label{sec:full}

The full PGE2 induced  cAMP signalling cascade for EP2 is described by the following system of ordinary differential equations combining the model for ligand-receptor binding \eqref{eq:lig_dyn} (with $m=1$)  with the models for the EP2 G-protein activation cycle  \eqref{EP2:model1} and cAMP expression \eqref{eq:EP2cAMP} (with $\eta=0$),
\begin{equation}\label{eq:EP2full}
\begin{aligned}
\frac{d p} {dt} & = - \mu_{1} \, r_1 \,   p^{n_1}  + \delta_1 r^\ast_1,  \qquad 
\frac{dr_1}{dt}  =   - \mu_{1} \, r_1  p^{n_1}  + \delta_1 r^\ast_1,  \qquad  
\frac{d r^\ast_1}{dt}  = \mu_{1} \, r_1 p^{n_1} - \delta_1 r_1^\ast,\\
\frac{d\alpha_s^\ast}{dt} &= k_{s}r^\ast_1 \frac{ b_1 - \beta}{b_1 - \beta + K_{r_1s}} - d_s \alpha^\ast_s,\\
 \frac{d\beta}{dt}& =k_{s}  r^\ast_1 \frac{ b_1 - \beta}{b_1 - \beta + K_{r_1s}} + d_{\beta s}(b_1 - \beta) - k_{\beta s} \beta(a_s - b_1 + \beta - \alpha^\ast_s),\\
  \frac{dw}{dt} & = k_{w}A^{tot}\frac{ \alpha^\ast_s }{\alpha^\ast_s+K_{as}}  - d_{w} P^{tot}\frac{ w}{w+K_{w}}. 
\end{aligned}
\end{equation}

The full PGE2 induced  cAMP signalling cascade for EP4 is described by the following system of ODEs combining the model for ligand-receptor binding~\eqref{eq:lig_dyn} (with $m=2$)  with the models for the EP4 G-protein activation cycles  \eqref{EP4:model3} and cAMP expression \eqref{eq:EP2cAMP} (with $\eta=1$),
\begin{align}\label{eq:EP4full}
\begin{split}
\frac{d p} {dt} & = - \mu_{2} \, r_2 \,   p  + \delta_2 r^\ast_2,   \qquad 
\frac{dr_2}{dt}  =   - \mu_{2} \, r_2 \, p  + \delta_2 r^\ast_2,    \qquad 
\frac{d r^\ast_2}{dt}  = \mu_{2} \, r_2 p - \delta_2 r_2^\ast-\rho_2r_2^*,\\
\frac{d\alpha_s^\ast}{dt} &= k_{s} r^*_2 \, f(\beta_i, \beta) - d_s   \alpha_s^\ast,  \qquad 
\frac{d\alpha_i^\ast}{dt} = \, k_{i} r^*_2  g(\beta_i, \beta)- d_i   \alpha_i^\ast,\\
\frac{d\beta_i}{dt}& =- k_{i} r^\ast_2 \, g(\beta_i, \beta) + k_{\beta i} \beta ( a_i - \beta_i - \alpha_i^\ast) - d_{\beta i} \beta_i, \\
\frac{d\beta}{dt} & = \;  k_{i} r^\ast_2\,   g(\beta_i, \beta)
 + k_{s} r^*_2  f(\beta_i, \beta) + d_{\beta i} \beta_i +  d_{\beta s} (b_2 - \beta- \beta_i)\\
& \qquad - \beta(k_{\beta s}(a_s - b_2 + \beta + \beta_i -\alpha_s^\ast) + k_{\beta i}(a_i - \beta_i - \alpha_i^\ast)),\\
  \frac{dw}{dt} & = k_{w}\frac{ \alpha^\ast_sA^{tot}}{\alpha^\ast_s + K_{as} (1+ \alpha^\ast_i /K_{ai})}  - d_{w} \frac{ wP^{tot}}{w+K_{w}}, 
\end{split}
\end{align}
where
$$
\begin{aligned} 
f(\beta_i, \beta) & = \frac{ b_2 - \beta- \beta_i}{b_2 - \beta  - \beta_i+ K_{r_2s} + (K_{r_2s}/K_{r_2i}) \beta_i},  \\
g(\beta_i, \beta) & = \frac{\beta_i}{\beta_i + K_{r_2i} + (b_2 - \beta - \beta_i)K_{r_2i}/K_{r_2s}} .
\end{aligned} 
$$

To describe crosstalk we combine the models for the single receptors as follows.
The combined activation of EP2 and EP4 by PGE2 is modeled by
\begin{equation}\label{eq:ligands_coupled}
\begin{aligned}
\frac{d p} {dt} & = - (1-\sigma) \mu_{1} \, r_1 \,   p^{n_1} -\sigma\mu_{2} \, r_2 \,   p  + \delta_1 r^\ast_1+ \delta_2 r^\ast_2,  \\
\frac{dr_1}{dt} & =   - (1-\sigma)\mu_{1} \, r_1  p^{n_1}  + \delta_{1} r^\ast_1, \qquad \quad 
\frac{d r^\ast_1}{dt}  =  (1-\sigma)\mu_{1} \, r_1 p^{n_1} - \delta_{1} r^\ast_1,\\
\frac{dr_2}{dt} & =   - \sigma\mu_{2} \, r_2  p  + \delta_{2} r^\ast_2, \qquad \qquad \qquad \;
\frac{d r^\ast_2}{dt}  =\sigma \mu_{2} \, r_2 p - \delta_{2} r^\ast_2 -\rho_{2}r^*_2,
\end{aligned}
\end{equation}
where $\sigma\in(0,1)$ indicates the fraction of PGE2 binding to EP4.
This system  is combined with the models for the G-protein activation cycles for EP2 \eqref{EP2:model1} and EP4 \eqref{EP4:model3} and cAMP expression \eqref{eq:EP2cAMP} (with $\eta=1$),
\begin{equation}\label{eq:EP2-EP4}
\begin{aligned}
\frac{d\alpha_s^\ast}{dt} &= \, k_{s} r^*_2 \sigma_0\, 
f(\beta_i, \beta)+k_{s}r^\ast_1 (1-\sigma_0)\frac{ b_2 - \beta-\beta_i}{b_2 - \beta -\beta_i+ K_{r_1s}} - d_s \,  \alpha_s^\ast,\\
\frac{d\alpha_i^\ast}{dt}&= \, k_{i} r^*_2\,  g(\beta_i, \beta) - d_i \,  \alpha_i^\ast,\\
\frac{d\beta_i}{dt}& =- k_{i} r^\ast_2 \, g(\beta_i, \beta)  + k_{\beta i} \beta ( a_i - \beta_i - \alpha_i^\ast) - d_{\beta i} \beta_i, \\
\frac{d\beta}{dt} & = \, k_{i} r^\ast_2 \, g(\beta_i, \beta)  
 + k_{s} r^*_2 \sigma_0 \; f(\beta_i, \beta) +k_{s}  r^\ast_1(1-\sigma_0) \frac{ b_2 - \beta-\beta_i}{b_2 - \beta -\beta_i+ K_{r_1s}} 
 + d_{\beta i} \beta_i \\
& \quad +  d_{\beta s} (b_2 - \beta- \beta_i) - \beta(k_{\beta s}(a_s - b_2 + \beta + \beta_i -\alpha_s^\ast) + k_{\beta i}(a_i - \beta_i - \alpha_i^\ast)),\\
 \frac{dw}{dt} & = k_{w}\frac{ \alpha^\ast_sA^{tot}}{\alpha^\ast_s + K_{as} (1+ \alpha^\ast_i /K_{ai})}  - d_{w} \frac{ wP^{tot}}{w+K_{w}}.
\end{aligned}
\end{equation}
where $\sigma_0\in(0,1)$ indicates the fraction of G$_{\alpha_s\beta\gamma}$-units binding to EP4.

\subsection{Experimentally observed cAMP response}\label{sec:sim_exp}

\begin{figure}[ht]
  \centering
    \includegraphics[height=125pt]{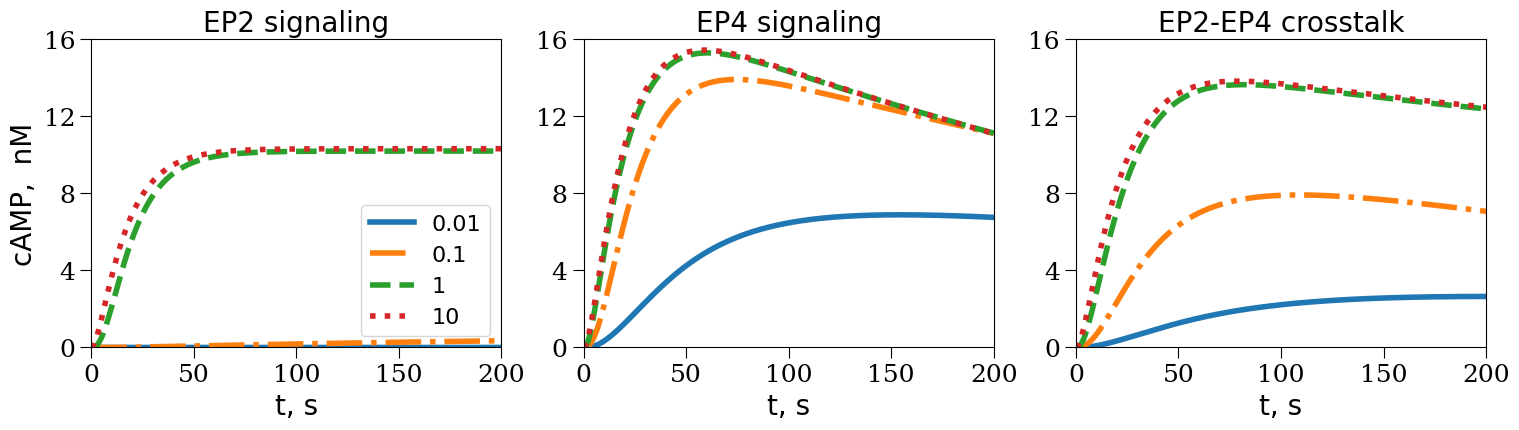}
    \caption{Simulated cAMP levels for the signaling pathways of EP2, EP4 and their crosstalk; models \eqref{eq:EP2full}--\eqref{eq:EP2-EP4} with parameter values in Tables~\ref{table:1}--\ref{table:3} for  PGE2 concentrations  $0.01\mu{\rm M}, 0.1\mu{\rm M}, 1\mu{\rm M}, 10\mu{\rm M}$}
    \label{im:cAMP_levels1}
\end{figure}

Figure \ref{im:cAMP_levels1} shows numerical simulations for the models \eqref{eq:EP2full}--\eqref{eq:EP2-EP4}. We only display cAMP levels for varying PGE2 concentrations to compare model predictions to the experimental data in Figures \ref{experiment} and \ref{im:schematic}. Simulation results for the complete signaling pathways for EP2, EP4 and their crosstalk are provided in subsequent subsections where the models' behavior and sensitivity to parameter variations is analyzed in greater detail.

Figure \ref{im:cAMP_levels1} shows that for the parameter values in Tables \ref{table:1}--\ref{table:3} the models \eqref{eq:EP2full} and \eqref{eq:EP4full} for the signaling pathways of EP2 and EP4 qualitatively predict the  cAMP response for the single receptors in Figures \ref{experiment} and \ref{im:schematic}. EP2 shows the characteristic threshold behavior with respect to PGE2 while cAMP levels for EP4 gradually increase with increasing PGE2 concentrations and then decay over time. However, the maximum cAMP level for EP4 is higher than for EP2. Moreover, while the gradual increase of cAMP levels with respect to PGE2  for crosstalk is correct, the maximum cAMP levels predicted by model \eqref{eq:ligands_coupled}--\eqref{eq:EP2-EP4} for the combined EP2--EP4 signaling are higher than for the signaling of the single receptor~EP2.

\begin{figure}[ht]
  \centering
    \includegraphics[height=125pt]{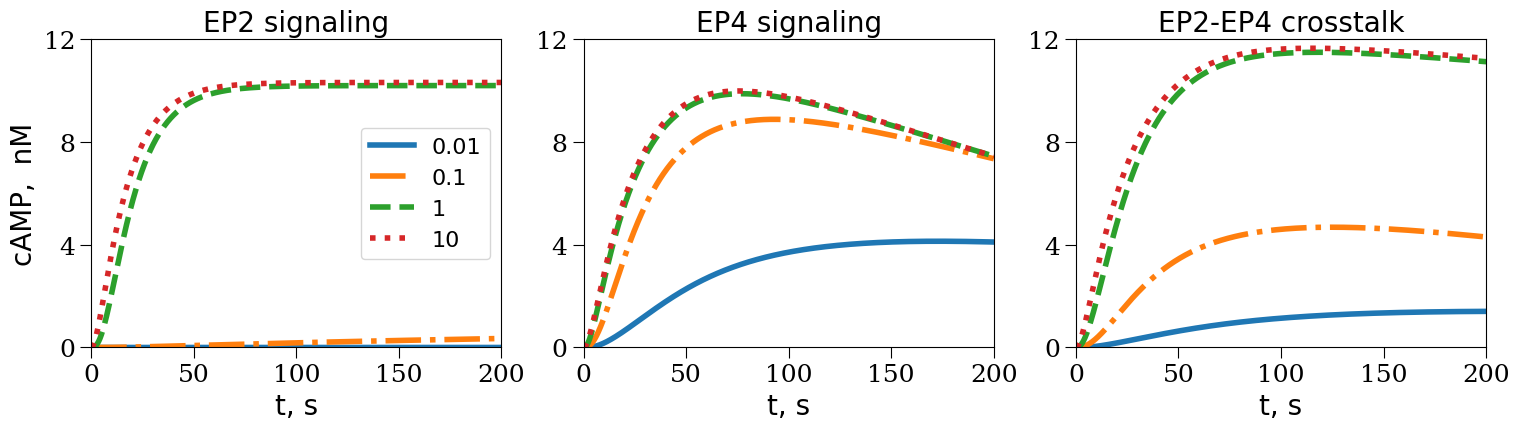}
    \caption{Simulated cAMP levels for the signaling pathways of EP2, EP4 and their crosstalk; models \eqref{eq:EP2full}--\eqref{eq:EP2-EP4} with parameter values in Tables~\ref{table:1}--\ref{table:3}, except $r_{2}^{tot}=4$nM, for  PGE2 concentrations  $0.01\mu{\rm M}, 0.1\mu{\rm M}, 1\mu{\rm M}, 10\mu{\rm M}$}
    \label{im:cAMP_levels2a}
\end{figure}

The total receptor concentrations $r_1^{tot}$ for EP2 and $r_2^{tot}$ for EP4 have an immediate impact on cAMP levels. Lowering $r_2^{tot}$ and assuming that $r_2^{tot}=r_1^{tot}$ the maximum cAMP levels for EP2 and EP4 coincide, as shown in Figure \ref{im:cAMP_levels2a}. 
As in Figure \ref{im:cAMP_levels1} we note that the models predict the qualitative cAMP response of the EP2 and EP4 correctly, but cAMP levels for the combined signaling of EP2 and EP4 are higher than for the signaling of the single receptors. 

\begin{figure}[ht]
  \centering
    \includegraphics[height=125pt]{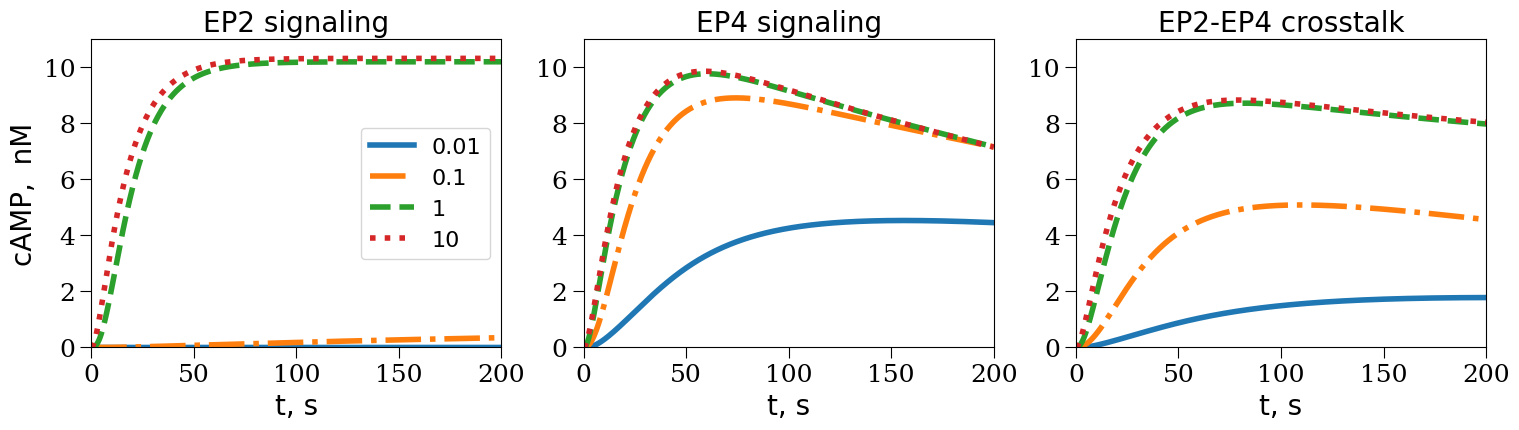}
    \caption{Simulated cAMP levels for the signaling pathways of EP2, EP4 and their crosstalk; models \eqref{eq:EP2full}--\eqref{eq:EP2-EP4} with parameter values in Tables~\ref{table:1}--\ref{table:3}, except $K_{ai}=0.017\mu M$, for  PGE2 concentrations  $0.01\mu{\rm M}, 0.1\mu{\rm M}, 1\mu{\rm M}, 10\mu{\rm M}$}
    \label{im:cAMP_levels2}
\end{figure}

Due to lacking information concerning the specific AC and PDE isoforms in the cAMP signaling pathways of EP2 and EP4 we use the simplistic model \eqref{eq:EP2cAMP} to describe cAMP regulation. It is very likely that relevant processes and/or the effects of additional enzymes that regulate subcellular cAMP levels are neglected. Even if the model is very simplistic, it suffices to modify $K_{ai}$, the dissociation constant for AC6 and G$_{\alpha i}$,  and use a slightly lower value to reproduce the correct qualitative cAMP response displayed in Figures \ref{experiment} and \ref{im:schematic}. Figure \ref{im:cAMP_levels2} shows that with  the parameter values in Tables \ref{table:1}--\ref{table:3}, except for $K_{ai}=0.017\mu M$, the models resemble the characteristic cAMP response for both, the single receptors and their crosstalk. The maximum cAMP levels for EP2 and EP4 coincide while the levels are lower for the combined signaling of the receptors.  
Hence, in the sequel we use $K_{ai}=0.017\mu M$  in the simulation studies, all other parameters are taken from Tables \ref{table:1}--\ref{table:3}.
Note that the parameter $K_{ai}$ is related to the G-protein G$_{\alpha_i}$ and hence, it only affects EP4 signaling and crosstalk but does not impact the EP2 signaling pathway.

\subsection{EP2 signaling pathway}\label{sec:sim_EP2}

\begin{figure}[ht]
  \centering
    \includegraphics[height=210pt]{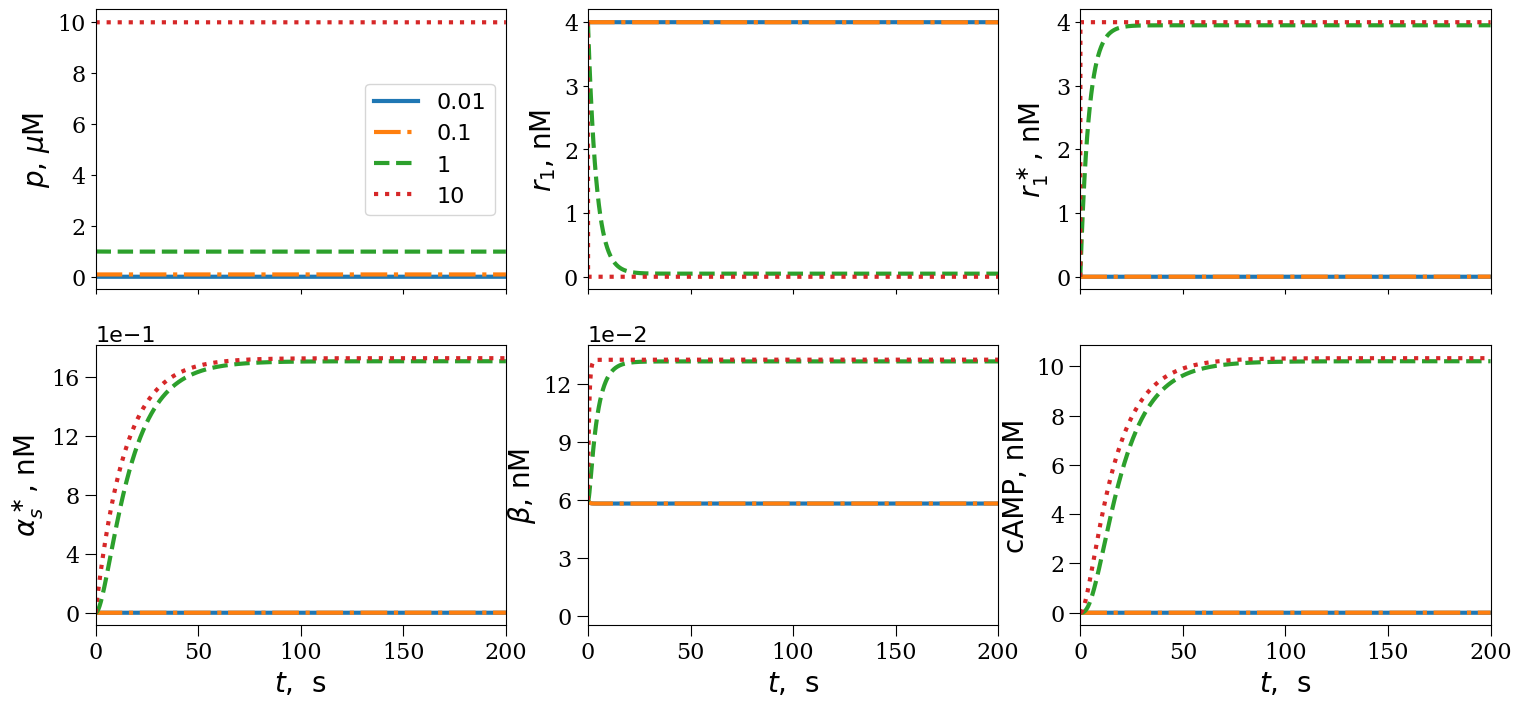}
    \caption{Simulation of the EP2 signaling pathway; model \eqref{eq:EP2full} with parameter values in Tables \ref{table:1}--\ref{table:3} for  PGE2 concentrations  $0.01\mu{\rm M}, 0.1\mu{\rm M}, 1\mu{\rm M}, 10\mu{\rm M}$}
    \label{im:EP2_standard}
\end{figure}

Figure \ref{im:EP2_standard} shows simulations for the full EP2 signaling cascade, i.e.~system \eqref{eq:EP2full} with  the parameter values in Tables \ref{table:1}--\ref{table:3}. The cAMP levels resemble the characteristic threshold behavior of EP2 in Figure \ref{experiment}. For PGE2 concentrations of $0.01\mu M$ and $0.1\mu M$ there is nearly no cAMP production while cAMP levels stabilize quickly for the PGE2 concentrations $1\mu M$ and $10\mu M$. 

\begin{figure}[ht]
  \centering
    \includegraphics[height=120pt]{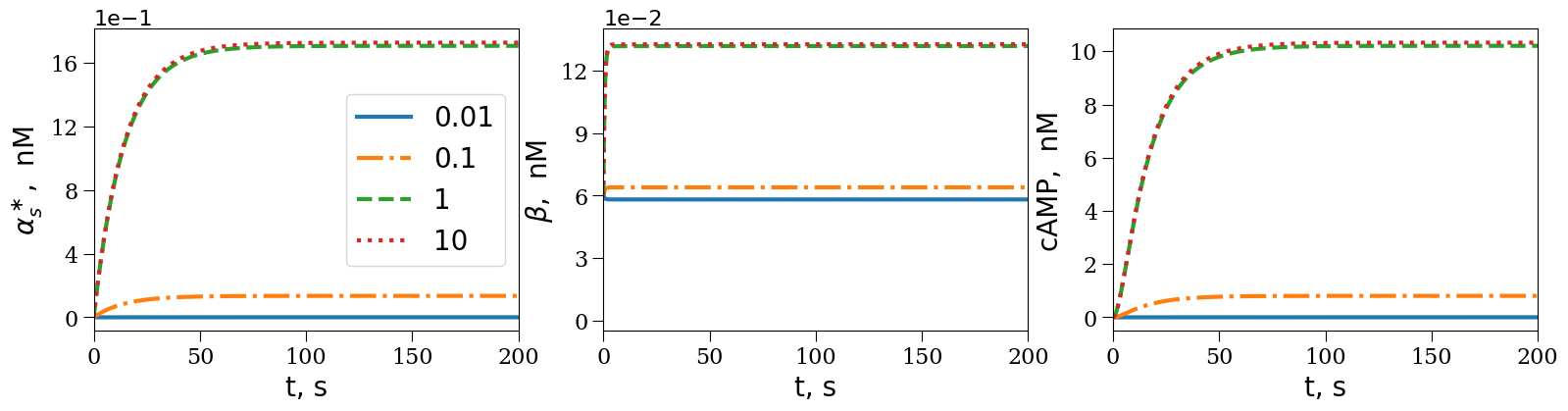}
    \caption{Simulation of the EP2 signaling pathway with  the quasi steady state approximation~\eqref{eq:rapidbinding} and parameter values in Tables \ref{table:1}--\ref{table:3} for  PGE2 concentrations  $0.01\mu{\rm M}, 0.1\mu{\rm M}, 1\mu{\rm M}, 10\mu{\rm M}$ }
    \label{im:EP2_without_ligands}
\end{figure}

We further observe that ligand receptor binding dynamics is rapid. Hence, we compare the results  with simulations  where the quasi steady state approximation \eqref{eq:rapidbinding} is used instead of the kinetic model \eqref{eq:lig_dyn} for ligand receptor binding. The simulations are shown in Figure \ref{im:EP2_without_ligands}. The models behave nearly the same  which justifies the quasi-steady state approximation. Both models predict the characteristic threshold behavior of EP2 in response to PGE2. 
However, in the sequel we include ligand receptor binding dynamics since it is important for EP4 and for modeling crosstalk of the receptors. 

\begin{figure}[ht]
  \centering
    \includegraphics[height=95pt]{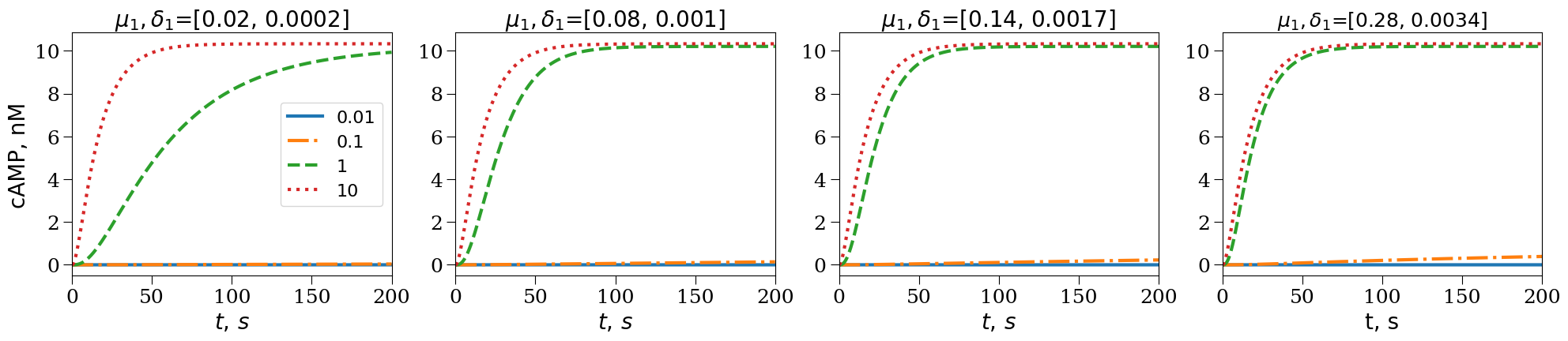}
    \caption{Simulated cAMP levels for the EP2 signaling pathway; model \eqref{eq:EP2full} with parameter values in Tables \ref{table:1}--\ref{table:3}  for varying  kinetic rate constants $\mu_1$ and $\delta_1$ keeping their ratio equal to $K_D=12$nM, for  PGE2 concentrations  $0.01\mu{\rm M}, 0.1\mu{\rm M}, 1\mu{\rm M}, 10\mu{\rm M}$}
    \label{im:EP2_kon_koff}
\end{figure}

The parameter values for the kinetic rate constants $\mu_1$ and $\delta_1$ for EP2 and PGE2 are not available in the literature. We therefore took the rate constant $\delta_1$ for the DcR3 decoy receptor in \cite{DeGhBa} and calculated $\mu_1$ given the equilibrium dissociation constant $K_D=12nM$ for PGE2 and EP2 in \cite{MaJaSoMl}. Figure \ref{im:EP2_kon_koff} confirms that varying the kinetic rate constants while keeping their ratio $\delta_1/\mu_1$ equal to the equilibrium dissociation constant $K_D$ has very little effect on cAMP levels for EP2.

\begin{figure}[ht]
  \centering
    \includegraphics[height=95pt]{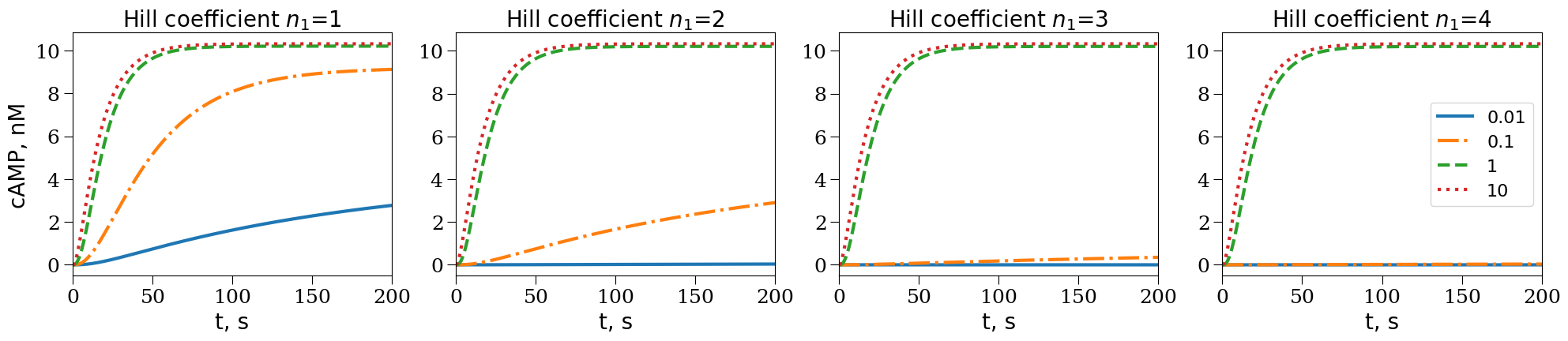}
    \caption{Simulated cAMP levels for the EP2 signaling pathway; model \eqref{eq:EP2full} with parameter values in Tables \ref{table:1}--\ref{table:3}  for varying  Hill coefficients, for  PGE2 concentrations  $0.01\mu{\rm M}, 0.1\mu{\rm M}, 1\mu{\rm M}, 10\mu{\rm M}$}
    \label{im:EP2_Hill_var}
\end{figure}

Finally, we analyze the impact of  the Hill coefficient $n_1$ in the EP2 ligand binding dynamics on cAMP response. Higher order reactions are crucial to describe the experimentally observed switch between low and high cAMP levels dependent on PGE2 concentrations. Figure \ref{im:EP2_Hill_var} shows cAMP levels predicted by model \eqref{eq:EP2full} with the parameter values in Tables \ref{table:1}--\ref{table:3} for varying Hill coefficients.
For the Hill coefficients $n_1=1$ and $n_1=2$ cAMP levels increase gradually with increasing PGE2 concentrations while the threshold behavior is modeled by Hill coefficients $n_1\geq 3$. 

\subsection{EP4 signaling pathway}\label{sec:sim_EP4}

\begin{figure}[ht]
  \centering
    \includegraphics[height=155pt]{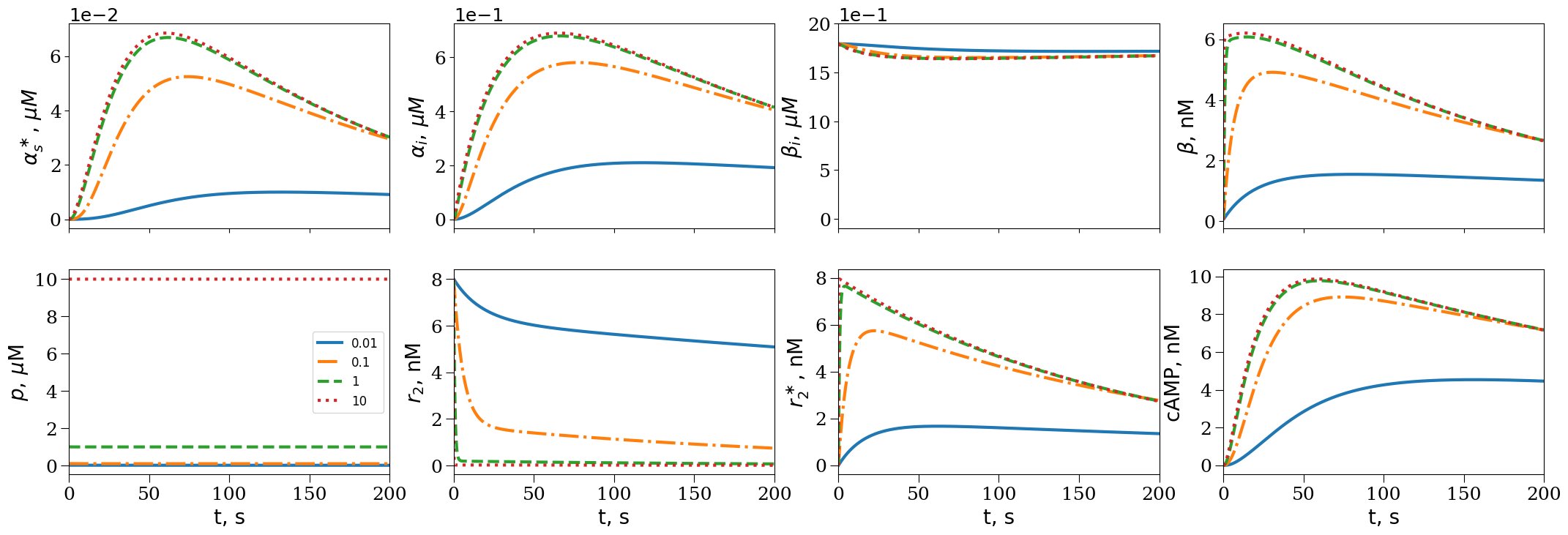}
    \caption{Simulation of the EP4 signaling pathway; model \eqref{eq:EP4full} with parameter values in Tables \ref{table:1}--\ref{table:3}, except $K_{ai}=0.017\mu M$, for  PGE2 concentrations  $0.01\mu{\rm M}, 0.1\mu{\rm M}, 1\mu{\rm M}, 10\mu{\rm M}$}
    \label{im:EP4-standard}
\end{figure}

Figure \ref{im:EP4-standard} shows simulations for the full EP4 signaling cascade, i.e.~model~\eqref{eq:EP4full} with the parameter values in Tables \ref{table:1}--\ref{table:3}, except $K_{ai}=0.017\mu M$. The cAMP levels resemble the characteristic signaling response of EP4 in Figure \ref{experiment}. The levels increase gradually with increasing PGE2 concentrations and show a sharp increase and then decay over time. 

\begin{figure}[ht]
  \centering
    \includegraphics[height=210pt]{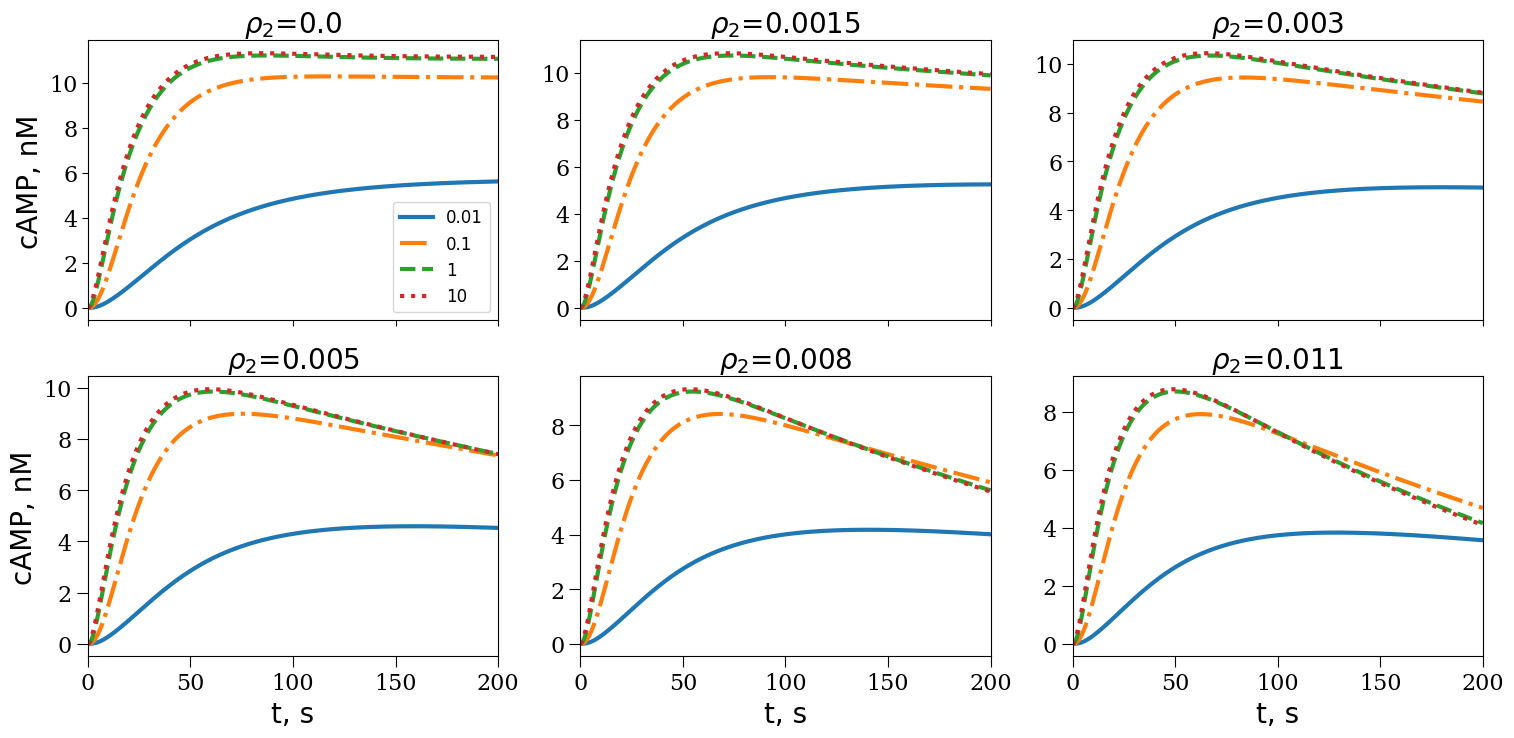}
    \caption{Simulated cAMP levels for the EP4 signaling pathway; model \eqref{eq:EP4full} with parameter values in Tables \ref{table:1}--\ref{table:3}, except $K_{ai}=0.017\mu M$,  for varying $\rho_2$ and  PGE2 concentrations  $0.01\mu{\rm M}, 0.1\mu{\rm M}, 1\mu{\rm M}, 10\mu{\rm M}$}
    \label{im:EP4_rho2_var}
\end{figure}

Different from EP2 where ligand receptor binding is rapid and the quasi-steady state assumption~\eqref{eq:rapidbinding} is justified, explicit modeling of ligand-receptor binding dynamics plays an important role for EP4 since the receptor internalizes rapidly. The kinetic model for  ligand-receptor interactions \eqref{eq:EP4full} provides a natural way to include internalization and the internalization  rate $\rho_2$  leads to the experimentally observed decay in cAMP levels over time. Figure~\ref{im:EP4_rho2_var} illustrates the effect of varying  $\rho_2$ on  the cAMP response of EP4. For $\rho_2=0$ the dynamics resembles the results for the EP4 signaling pathway with the quasi steady state approximation, as in~\eqref{eq:rapidbinding} but with $r_2^\ast$ and $r_2^{tot}$ and the corresponding parameters, see Figure~\ref{im:EP4_without_ligands} in Appendix. We note that altering the internalization rate changes decay properties but has little impact on maximum cAMP levels.

\begin{figure}[ht]
  \centering
    \includegraphics[height=210pt]{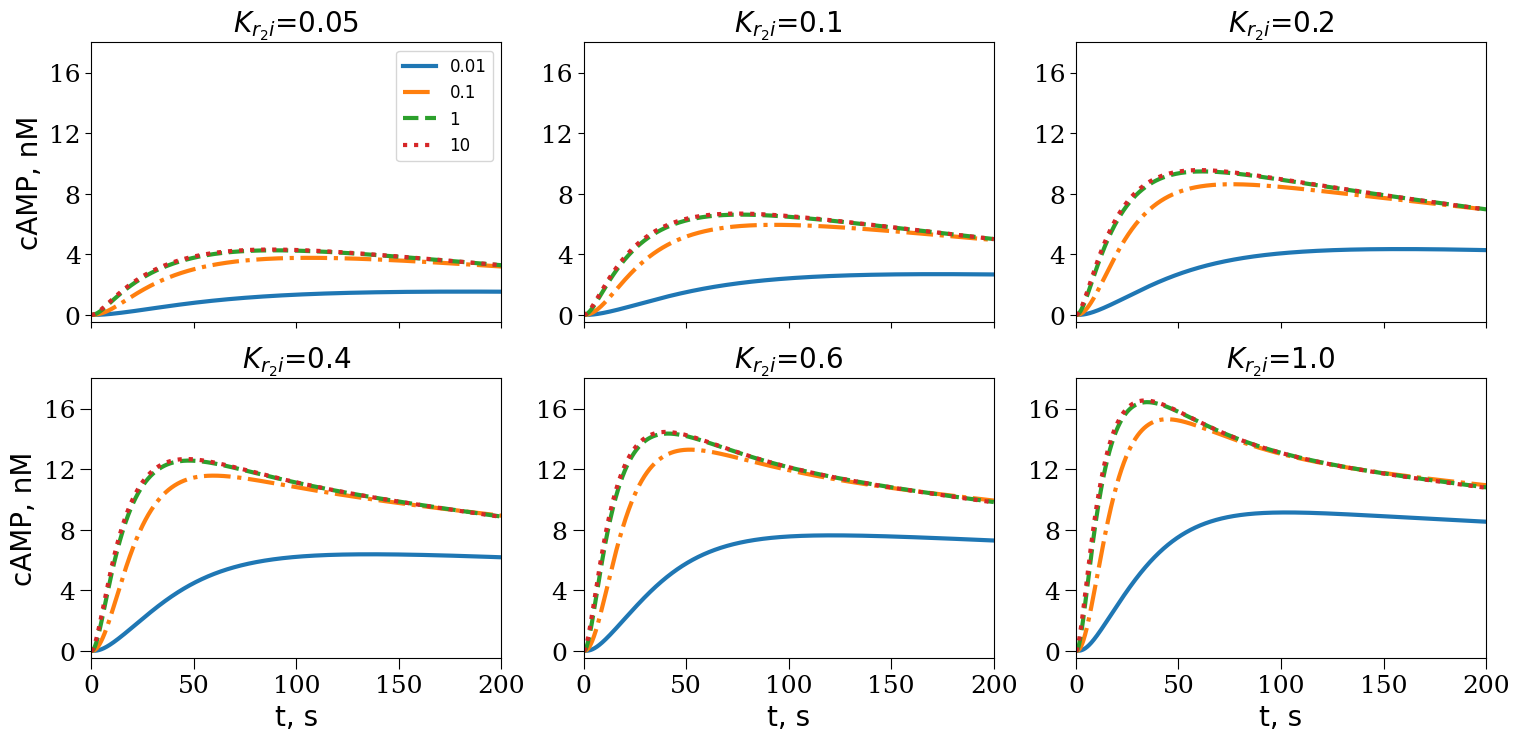}
    \caption{Simulated cAMP levels for the EP4 signaling pathway; model \eqref{eq:EP4full} with parameter values in Tables \ref{table:1}--\ref{table:3}, except $K_{ai}=0.017\mu M$,  for varying $K_{r_2i}$ and  PGE2 concentrations  $0.01\mu{\rm M}, 0.1\mu{\rm M}, 1\mu{\rm M}, 10\mu{\rm M}$}
    \label{im:EP4_Kr2i_var}
\end{figure}

EP4 activates the stimulating G-protein G$_{\alpha_s}$ and the inhibiting G-protein G$_{\alpha_i}$.  In model~\eqref{eq:EP4full} the effects of competitive binding of G$_{\alpha_s}$ and G$_{\alpha_i}$ to the receptor are determined by the ratio  $K_{r_2s}/K_{r_2i}$ of the EP4-dissociation rates for G$_{\alpha_s}$ and G$_{\alpha_i}$.  Figure~\ref{im:EP4_Kr2i_var} shows the impact of competitive binding on the EP4 signaling pathway where cAMP levels for different values of $K_{r_2i}$ are shown. Higher values for $K_{r_2i}$ lead to higher cAMP levels. Similarly, varying $K_{r_2s}$ has an opposite effect on cAMP response, i.e.~cAMP levels decrease with increasing $K_{r_2s}$, see Figure~\ref{im:EP4_Kr2s_var} in Appendix.

Competitive binding also plays an important role for cAMP expression since the inhibiting G-protein $G_{\alpha_i}$ prevents binding of AC to the stimulating G-protein $G_{\alpha_s}$. In  model~\eqref{eq:EP4full} competitive binding of $G_{\alpha_s}$ and $G_{\alpha_i}$ to AC  is  determined by the parameter $K_{ai}$. 
Varying the dissociation constant $K_{ai}$ for AC6 and G$_{\alpha_i}$  only affects cAMP levels but does not change the qualitative cAMP dynamics nor the gradual increase of cAMP  with respect to PGE2, see Figure~\ref{im:EP4_Kai_var} in Appendix.

\subsection{EP2-EP4  crosstalk}\label{sec:sim_EP2EP4}

\begin{figure}[ht]
  \centering
    \includegraphics[height=280pt]{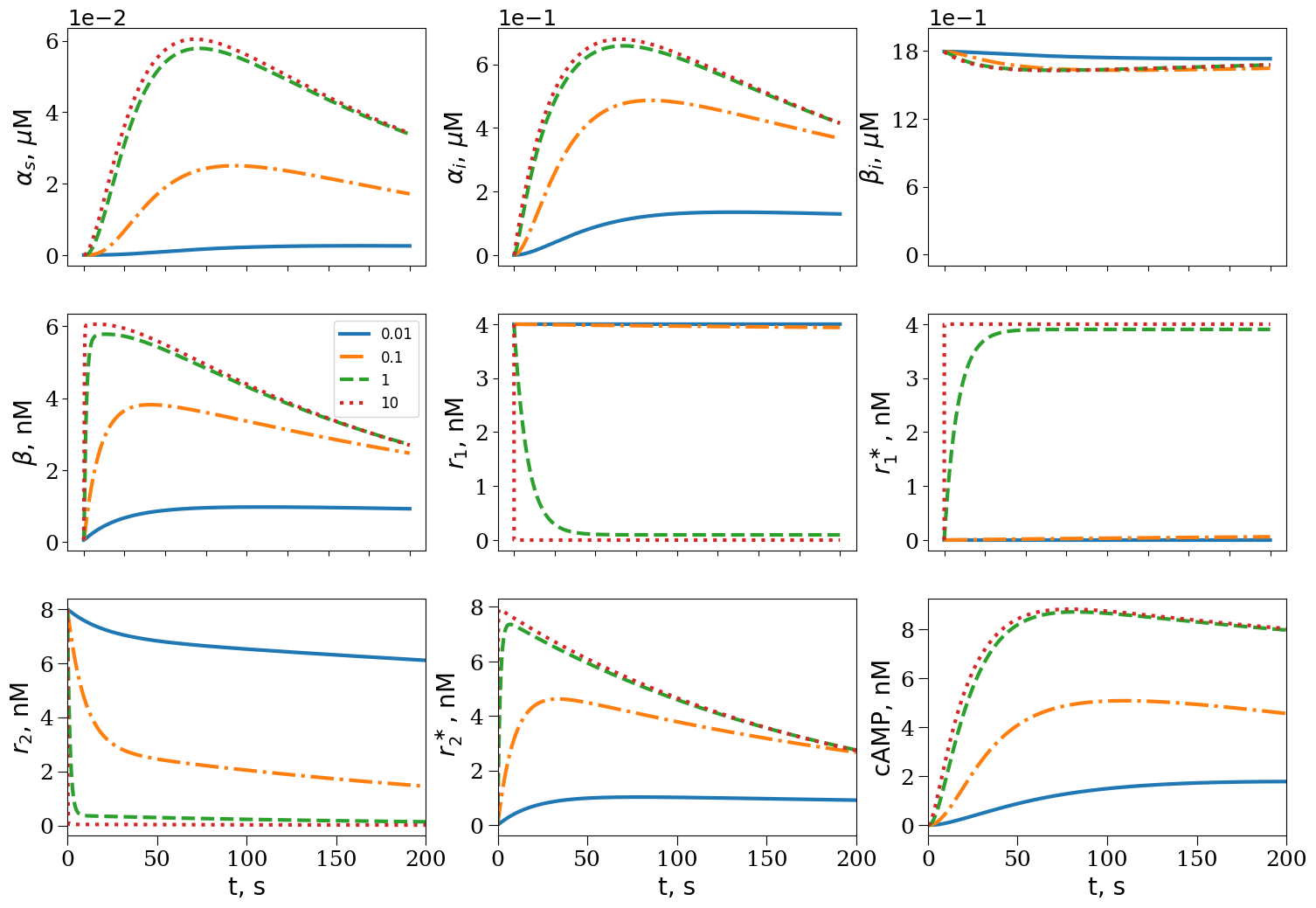}
    \caption{Simulation of the combined EP2-EP4 signaling pathway; model \eqref{eq:ligands_coupled}--\eqref{eq:EP2-EP4}  with parameter values in Tables \ref{table:1}--\ref{table:3}, except $K_{ai}=0.017\mu M$, for  PGE2 concentrations  $0.01\mu{\rm M}, 0.1\mu{\rm M}, 1\mu{\rm M}, 10\mu{\rm M}$}
    \label{im:EP2_EP4_standard}
\end{figure}

Figure \ref{im:EP2_EP4_standard} shows simulations for the combined signaling pathways of EP2 and EP4, i.e.\ model~\eqref{eq:ligands_coupled}--\eqref{eq:EP2-EP4} with the parameter values in Tables \ref{table:1}--\ref{table:3}, except $K_{ai}=0.017\mu M$. The cAMP levels resemble the characteristic signaling response for receptor crosstalk in Figure \ref{experiment}. As for the receptor EP4 the cAMP levels increase gradually with increasing PGE2 concentrations and show a sharp increase and then decay over time. However, the maximum cAMP levels are lower than for the signaling of EP2 and EP4 alone, see also Figure \ref{im:cAMP_levels2}.

The impact of varying the EP4 internalization rate $\rho_2$ on EP2-EP4 crosstalk is similar as for EP4. Increasing $\rho_2$ leads to a larger decay in cAMP concentration over time and for $\rho_2=0$ the dynamics resembles the model without ligand dynamics, i.e.\ replacing the equations for ligand-binding dynamics by the quasi steady state approximation, see Figure~\ref{im:EP2_EP4_rho2_var_1} in Appendix.
The effects of competitive binding in the G-protein activation cycles of EP4 are determined by the ratio  $K_{r_2s}/K_{r_2i}$ of the  dissociation rates for for G$_{\alpha_i}$ and G$_{\alpha_s}$. Varying the dissociation rate  $K_{r_2i}$ has the same qualitative effect on cAMP levels for crosstalk  as for single receptor EP4 shown in Figure~\ref{im:EP4_Kr2i_var},   see Figure~\ref{im:EP2_EP4_Kr2i_var} in Appendix.
The inhibiting effect of G$_{\alpha_i}$ on cAMP production is modeled by competitive binding between G$_{\alpha_i}$ and G$_{\alpha_s}$ to AC and determined by the dissociation constant $K_{ai}$ for AC6 and G$_{\alpha_i}$ in \eqref{eq:EP2-EP4}. 
Using a slightly lower value, $K_{ai}=0.017\mu M$, than in Table~\ref{table:3} was necessary to obtain the correct cAMP response for crosstalk, see Figure~\ref{im:cAMP_levels1}.  Increasing $K_{ai}$ leads to higher cAMP levels for crosstalk, but does not change the qualitative cAMP dynamics in response to PGE2,   see Figure~\ref{im:EP2_EP4_Kai_var} in Appendix. Similarly to EP4 signaling, cAMP levels decrease with increasing $K_{r_2s}$, see Figure~\ref{im:EP2_EP4_Kr2s_var} in Appendix.

\begin{figure}[ht]
  \centering
    \includegraphics[height=200pt]{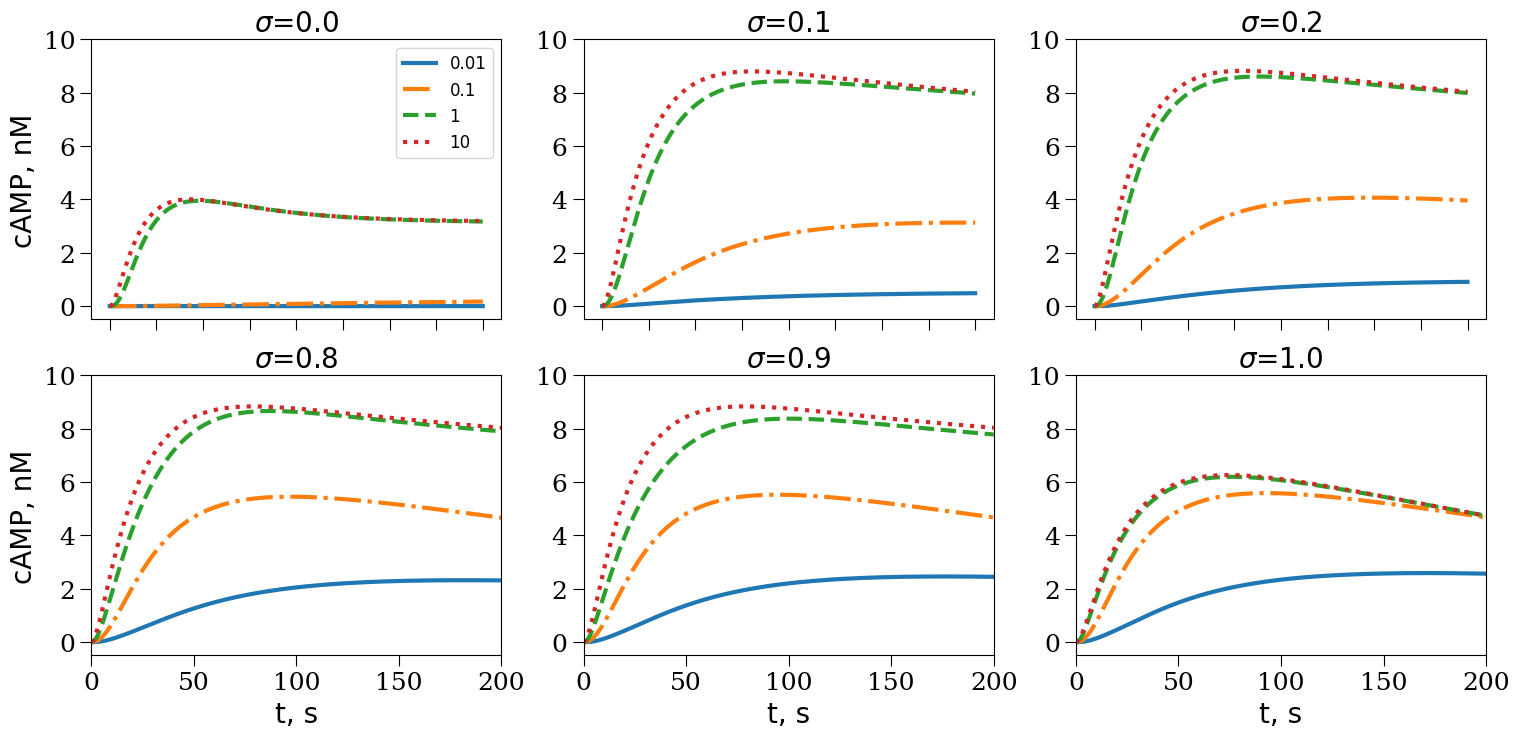}
    \caption{Simulated cAMP levels for the combined EP2-EP4 signaling pathway; model \eqref{eq:ligands_coupled}--\eqref{eq:EP2-EP4}  with parameter values in Tables \ref{table:1}--\ref{table:3}, except $K_{ai}=0.017\mu M$, for varying $\sigma$ and PGE2 concentrations  $0.01\mu{\rm M}, 0.1\mu{\rm M}, 1\mu{\rm M}, 10\mu{\rm M}$}
    \label{im:EP2_EP4_sigma_var}
\end{figure}

Finally, we analyze the effects of the crosstalk binding ratios  $\sigma,\sigma_0\in(0,1)$ in model \eqref{eq:ligands_coupled}--\eqref{eq:EP2-EP4}. Figure \ref{im:EP2_EP4_sigma_var} shows cAMP levels when varying  $\sigma$. The binding ratio $\sigma$ indicates the fraction of PGE2 binding to EP4 while $1-\sigma$ is the fraction of PGE2 binding to EP2. 
For $\sigma=0$ the cAMP levels resemble the results for EP2 in Figure~\ref{im:cAMP_levels2} while for $\sigma=1$ the cAMP levels resemble the results for EP4 in Figure~\ref{im:cAMP_levels2}. Note that cAMP levels for $\sigma=0$ and $\sigma=1$ are lower than for the single receptors, see Figure~\ref{im:cAMP_levels2}, due to the binding ratio $\sigma_0=0.5$ in the G-protein activation cycles indicating that half of the G$_{\alpha_s}$ protein complexes bind to  EP2 and half of them to EP4. 

\begin{figure}[ht]
  \centering
    \includegraphics[height=200pt]{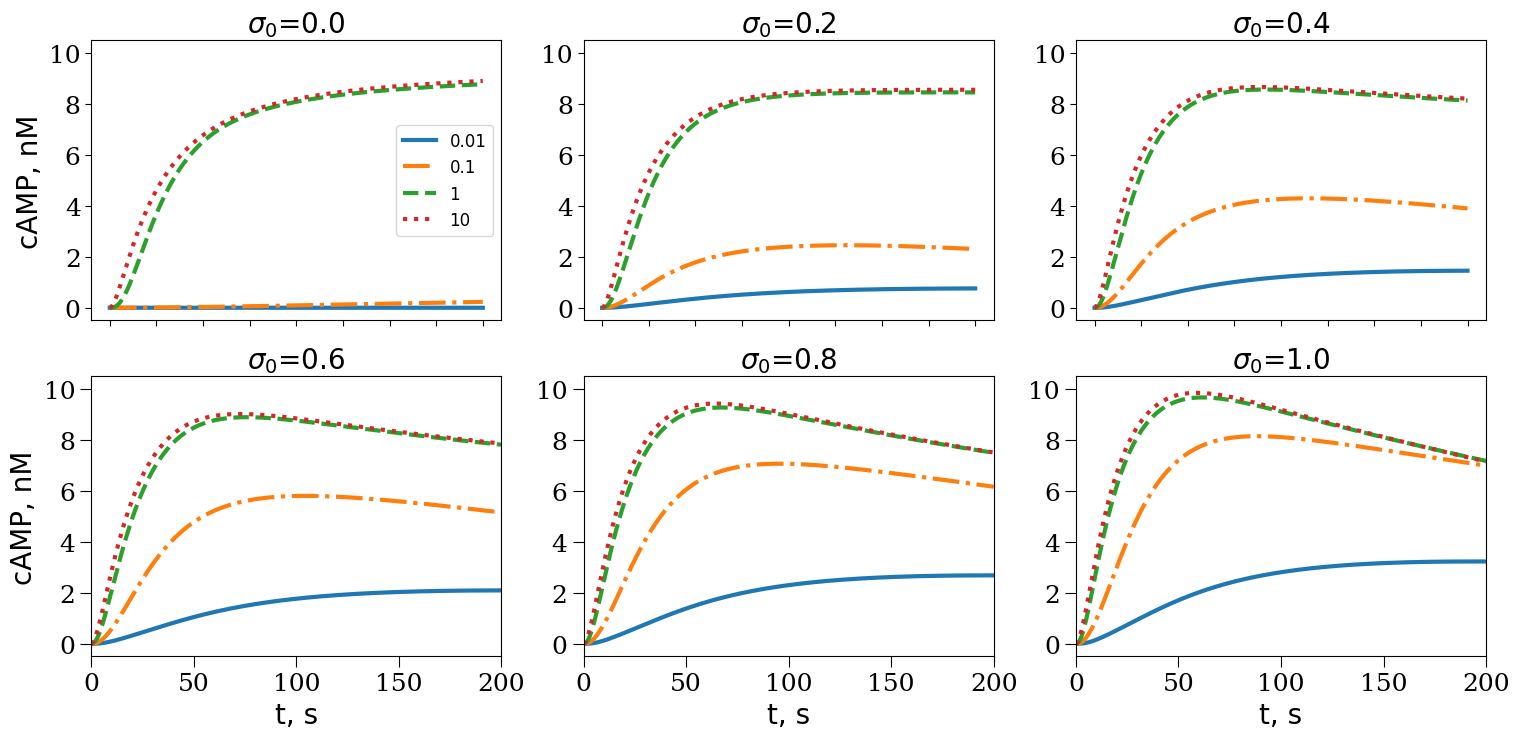}
    \caption{Simulated cAMP levels for the combined EP2-EP4 signaling pathway; model \eqref{eq:ligands_coupled}--\eqref{eq:EP2-EP4}  with parameter values in Tables \ref{table:1}--\ref{table:3}, except $K_{ai}=0.017\mu M$, for varying $\sigma$ and PGE2 concentrations  $0.01\mu{\rm M}, 0.1\mu{\rm M}, 1\mu{\rm M}, 10\mu{\rm M}$}
    \label{im:EP2_EP4_sigma0_var}
\end{figure}

Figure~\ref{im:EP2_EP4_sigma0_var} shows cAMP levels when varying the binding ratio  $\sigma_0$ for EP2 and EP4 and the G-protein G$_{\alpha_s}.$ Small values of $\sigma_0$ reflect that G$_{\alpha_s}$ binds with high probability to EP2 and with low probability to EP4. For $\sigma_0=0$ the cAMP levels for crosstalk  resemble the results for EP2 signaling while for $\sigma_0=1$  the results for EP4 signaling, see  Figure~\ref{im:cAMP_levels2}. Simultaneous variations in $\sigma$ and $\sigma_0$, with $\sigma= \sigma_0$,  affect the qualitative cAMP dynamics, where the cAMP levels for lower values  resemble the results for EP2 and for higher values the results for EP4 signaling,   but have smaller impact on the maximum levels of cAMP when $\sigma=\sigma_0=0$ and $\sigma=\sigma_0=1$, see Figure~\ref{im:EP2_EP4_sigma_sigma_0_var} in Appendix.

\section{Conclusion and discussion}\label{sec:con}

 We propose mathematical models for the cAMP signaling cascades of the GPCRs EP2 and EP4 and their crosstalk. The models qualitatively predict the characteristic signaling behavior of the receptors and their crosstalk in response to varying PGE2 concentrations as observed   experimentally \cite{Vleeshouwers}. The models  allow us to identify processes in the complex signaling pathways that can explain the differences in the cAMP profiles for EP2 and EP4 and their combined signaling. 
Crucial for the different behavior seem to be processes in the ligand binding kinetics of the receptors. 
While many models assume ligand-receptor binding is rapid and make quasi steady state assumptions, it is becoming increasingly important to incorporate specific ligand-receptor binding kinetics and processes such as internalization \cite{Carvalho, Bridge}.

Ligand binding dynamics played a crucial role in our models, allow us to explain the qualitative differences in the cAMP profiles of EP2 and EP4.
Hence, different from most models for GPCRs  we include a dynamic model for ligand-receptor binding where we take receptor specific behavior into account such as ultrasensitivity of EP2 and internalization for EP4. 
Ligand dynamics played a crucial role in our models, allowing for us on one hand  explain the qualitative differences in the cAMP profiles of EP2 and EP4, and on the other hand to model crosstalk.
While several models exist that describe crosstalk of receptors in response to different ligands \cite{Leander} we are not aware of models for the crosstalk of receptors that bind to the same ligand as considered here. Another novelty in our work concerns the model for EP4 as this receptor activates two different G-proteins. We model the EP4 activation cycles for G$_{\alpha_s}$ and G$_{\alpha_i}$ and include the effects of competitive binding. Models for the G-protein activation cycles of GPCRs typically assume that only one G-protein is activated \cite{Bridge,Katanaev,Leander, Woodroffe}.

In the ligand-receptor binding kinetics ultrasensitivity of EP2 with respect to PGE2 is modeled by a Hill coefficient $n_1=3$ which leads to the threshold behavior shown in Figure~\ref{experiment}. Another possible explanation for the threshold behavior in EP2 signaling is bistability \cite{Roth}, i.e.\ the system can switch between two distinct stable steady states \cite{Markevich}. Bistability can be approximated by a large Hill coefficient $n_1$, but further experimental data is required to determine whether EP2 shows bistability or ultrasensitivity. 
A possible explanation for ultrasensitivity is dimerization. If binding of PGE2 to EP2 is followed by the formation of dimers this can increase  the probability that another receptor starts signaling
and hence, leads to positive cooperativity \cite{Bouhaddou}. It was shown
that dimerization can dominate over the canonical allosteric mechanism for generating cooperativity~\cite{Bouhaddou}, i.e.\ it can result in positive cooperativity and thus, in a Hill coefficient  greater than one. However, while dimerization is documented for other GPCRs it remains to be identified for EP2 \cite{Vleeshouwers}. 
Here we choose the Hill coefficient $n_1=3$ since Hill coefficients greater than three are considered very large when modeling dimerization. 
Different from EP2 the cAMP expression for EP4 increases gradually with PGE2 levels which is modeled by the Hill coefficient $n_2=1$.
While EP2 does not internalize and ligand receptor binding is rapid, internalization plays a key role for EP4. Internalization is taken into account in the kinetic model for ligand receptor binding and leads to a peak and then the decay in cAMP levels over time for EP4 as displayed in Figure \ref{experiment}.

Due to too many unknown aspects concerning the specific AC and PDE isoforms, and possibly additional processes involved in intracellular cAMP regulation, we use a simplistic model for cAMP production assuming 
that AC6 and PDE4 are the dominant isoforms regulating  cAMP levels. However, other isoforms can  be involved, instead of, or in addition to, AC6 and PDE4. The regulation patterns for AC isoforms differ significantly, see Table~\ref{table:1}, and some isoforms are also inhibited or stimulated by the G-protein G$_{\beta\gamma}$ which we do not take into account here. For an improved model, additional knowledge on the specific AC and PDE isoforms is required. Our model provides a basis and framework for future studies. It allows to identify key parameters responsible for cAMP regulation in the EP2 and EP4 signaling pathways. The simulation results illustrate how changes in parameter values affect the EP2 and EP4 cAMP response and their crosstalk. 
One could model and examine the effects of  different AC isoforms on cAMP levels by modifying the equation for cAMP production~\eqref{eq:EP2cAMP}. 

Many aspects in the cAMP signaling cascades of EP2 and EP4 are still unknown.  Moreover, few precise parameter values are available for the specific cell types and GPCRs used for the experimental data displayed in Figure \ref{experiment}. 
Some of the parameter values in the simulations stem from different organisms, cell types and/or receptors and typically, values vary significantly depending on the cell type, organism and experimental setup.
Here, we balance between detailed modeling of  the important steps in the cAMP signaling pathways of EP2 and EP4 and inserting uncertainties due to unknown parameter values.
Finally, we remark that the experimental data only indicates relative cAMP levels, but do not provide information on the order of magnitude of intracellular cAMP levels. Hence,
our models focus on qualitative predictions, but do not provide  quantitative outcomes. 

Increased PGE2 concentrations are presented in the tumor microenvironment of several cancer types \cite{Kobayashi}. Since PGE2 regulates immune cell function, the selective modulation of EP receptor signaling pathways has been shown to improve antitumor immune responses \cite{Majumder}. Better insight into the EP2 and EP4 signaling behaviors is crucial to efficiently control the cellular responses to PGE2. The mathematical model developed here could contribute to the design of receptor specific antagonists to be used as anticancer therapies.

\section*{Acknowledgements}
D. Lidke acknowledges funding from the National Institutes of Health, grant R35GM126934.

\bibliographystyle{acm}
\bibliography{sample}

\renewcommand{\thefigure}{A\arabic{figure}}
\renewcommand{\theequation}{A\arabic{equation}}
\setcounter{figure}{0}
\setcounter{equation}{0}

\section*{Appendix: Mathematical model for EP2 and EP4 signaling pathways and their crosstalk }\label{models_S}

Quasi steady state approximation for ligand-receptors dynamics
\begin{equation}\label{Seq:quasiEP4}
 r_2^\ast= \frac{r_2^{tot} p}{K_D + p}, \qquad  \text{ where } \; K_D = \frac{\delta_2}{\mu_2}.
\end{equation}

\begin{figure}[ht]
  \centering
    \includegraphics[height=95pt]{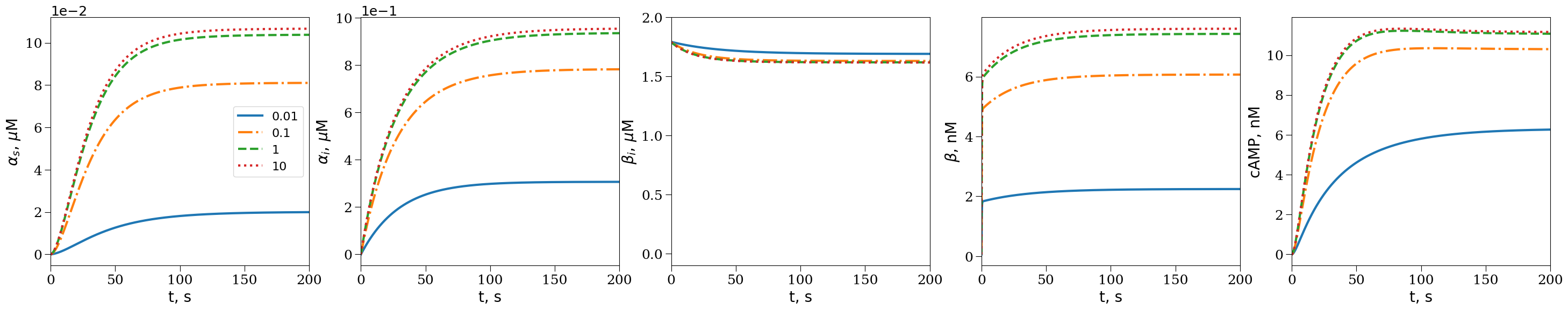}
    \caption{Simulation of the EP4 signaling pathway with the quasi steady state approximation; model~\eqref{eq:EP4full} with ligand-receptors dynamics replaced by \eqref{Seq:quasiEP4}, with parameter values in Tables~2--3, except $K_{ai}=0.017\mu M$, for  PGE2 concentrations  $0.01\mu{\rm M}, 0.1\mu{\rm M}, 1\mu{\rm M}, 10\mu{\rm M}$}
    \label{im:EP4_without_ligands}
\end{figure}

\begin{figure}[ht]
  \centering
    \includegraphics[height=105pt]{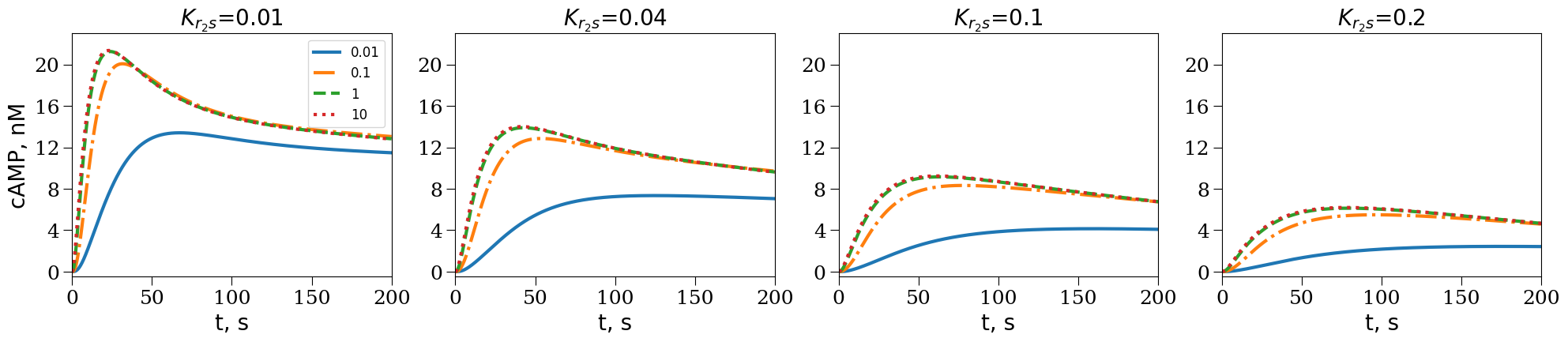}
    \caption{Simulated cAMP levels for the EP4 signaling pathway; model~\eqref{eq:EP4full} with parameter values in Tables~2--4, except $K_{ai}=0.017\mu M$,  for varying $K_{r_2s}$ and  PGE2 concentrations  $0.01\mu{\rm M}, 0.1\mu{\rm M}, 1\mu{\rm M}, 10\mu{\rm M}$ }
    \label{im:EP4_Kr2s_var}
\end{figure}

\begin{figure}[ht]
  \centering
    \includegraphics[height=105pt]{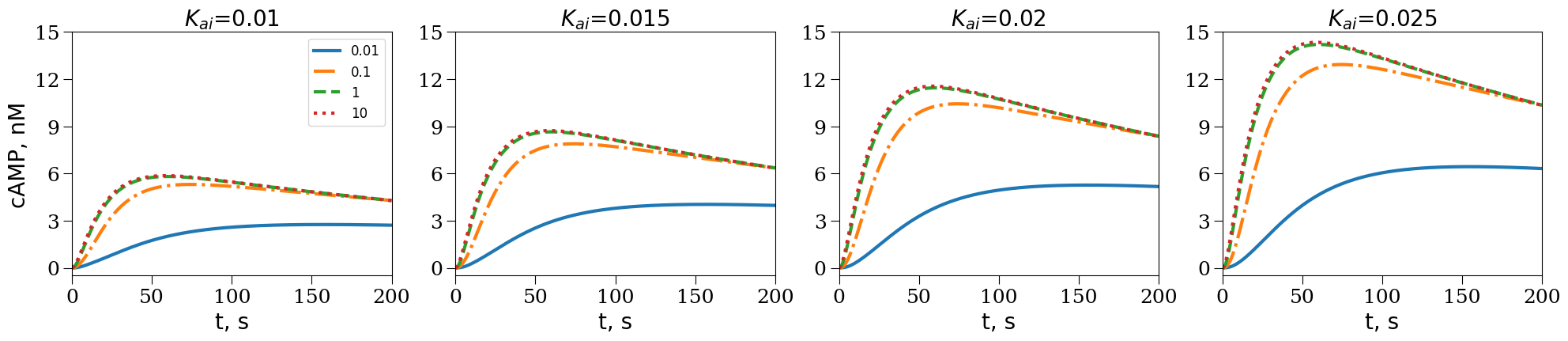}
    \caption{Simulated cAMP levels for the EP4 signaling pathway; model~\eqref{eq:EP4full} with parameter values in Tables~2--4,   for varying $K_{ai}$ and  PGE2 concentrations  $0.01\mu{\rm M}, 0.1\mu{\rm M}, 1\mu{\rm M}, 10\mu{\rm M}$}
    \label{im:EP4_Kai_var}
\end{figure}

\begin{figure}[ht]
  \centering
    \includegraphics[height=220pt]{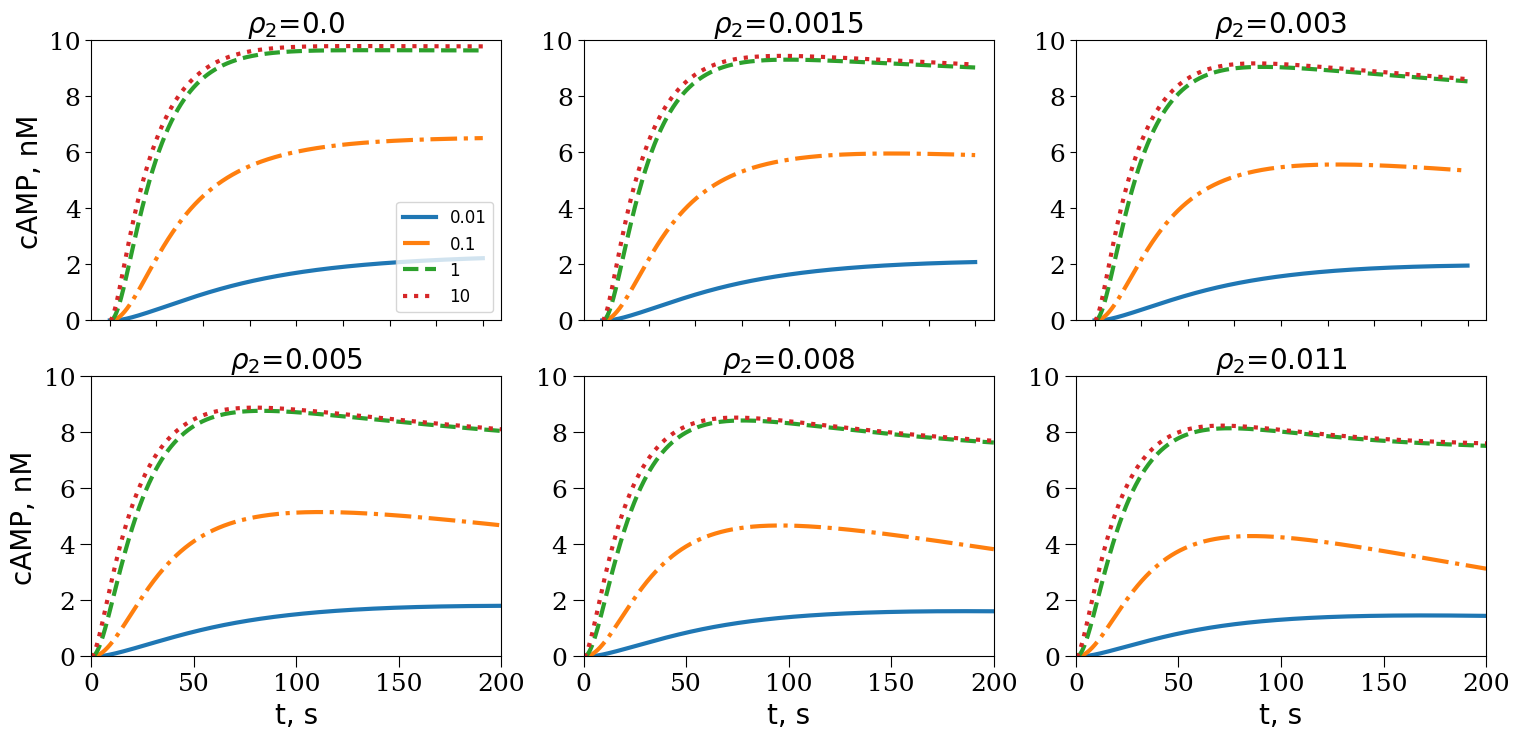}
    \caption{Simulated cAMP levels for the combined EP2-EP4 signaling pathway; model~\eqref{eq:ligands_coupled}--\eqref{eq:EP2-EP4}  with parameter values in Tables~2--4, except $K_{ai}=0.017\mu M$, for varying $\rho_2$ and PGE2 concentrations  $0.01\mu{\rm M}, 0.1\mu{\rm M}, 1\mu{\rm M}, 10\mu{\rm M}$}
    \label{im:EP2_EP4_rho2_var_1}
\end{figure}

\begin{figure}[ht]
  \centering
    \includegraphics[height=220pt]{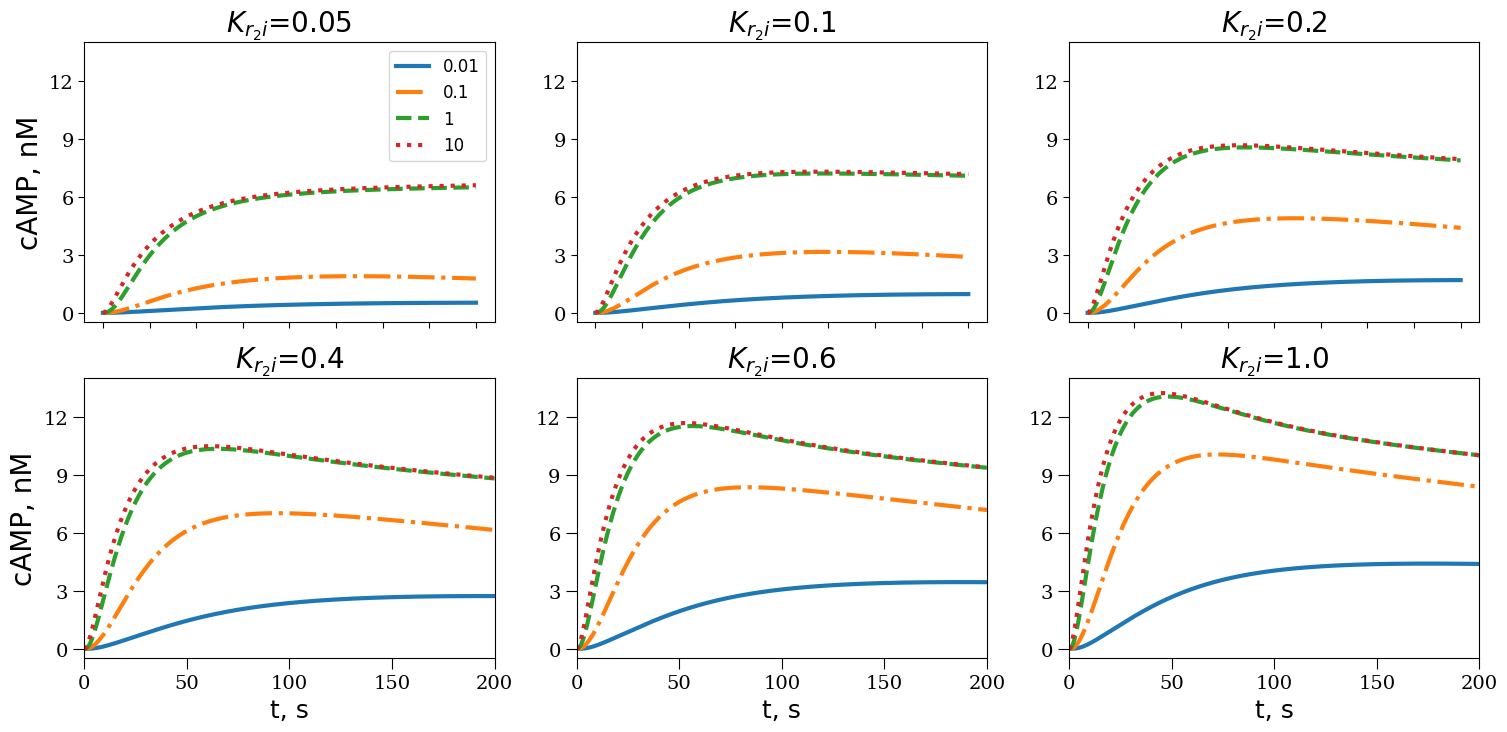}
    \caption{Simulated cAMP levels for the combined EP2-EP4 signaling pathway; model~\eqref{eq:ligands_coupled}--\eqref{eq:EP2-EP4}  with parameter values in Tables~2--4, except $K_{ai}=0.017\mu M$, for varying $K_{r_2i}$ and PGE2 concentrations  $0.01\mu{\rm M}, 0.1\mu{\rm M}, 1\mu{\rm M}, 10\mu{\rm M}$}
    \label{im:EP2_EP4_Kr2i_var}
\end{figure}

\begin{figure}[ht]
  \centering
    \includegraphics[height=105pt]{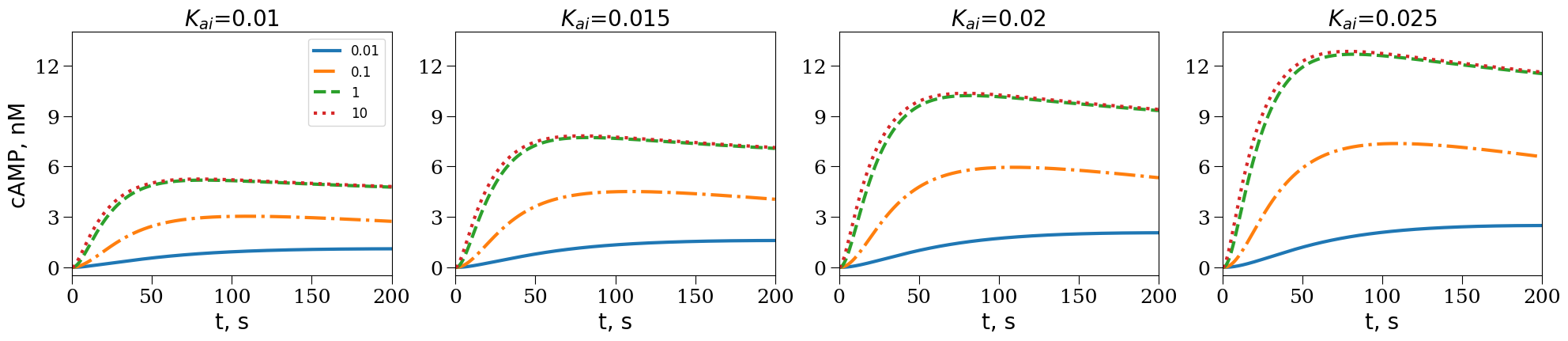}
    \caption{Simulated cAMP levels for the combined EP2-EP4 signaling pathway; model~\eqref{eq:ligands_coupled}--\eqref{eq:EP2-EP4}  with parameter values in Tables~2--4 for varying $K_{ai}$ and PGE2 concentrations  $0.01\mu{\rm M}, 0.1\mu{\rm M}, 1\mu{\rm M}, 10\mu{\rm M}$}
    \label{im:EP2_EP4_Kai_var}
\end{figure}

\begin{figure}[ht]
  \centering
    \includegraphics[height=105pt]{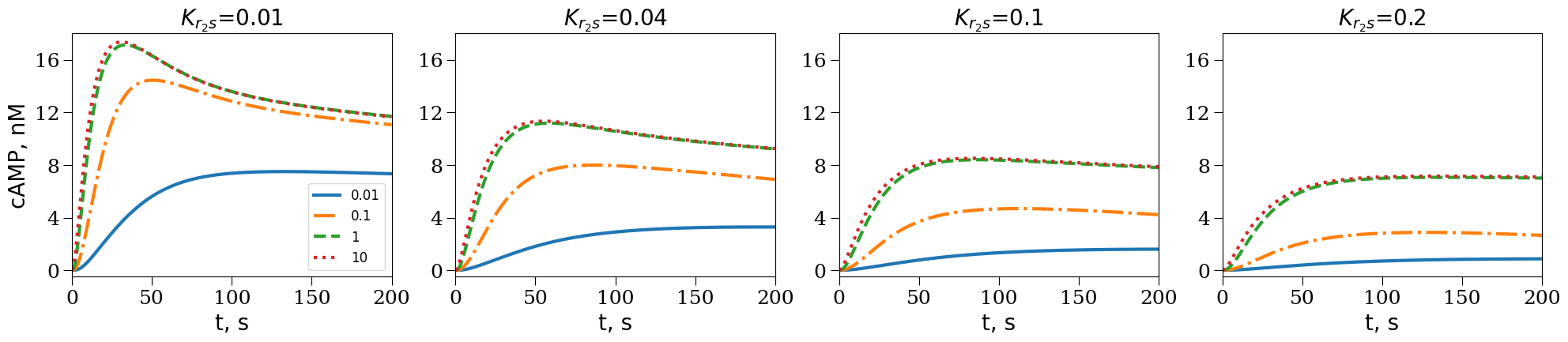}
    \caption{Simulated cAMP levels for the combined EP2-EP4 signaling pathway; model~\eqref{eq:ligands_coupled}--\eqref{eq:EP2-EP4}  with parameter values in Tables~2--4, except $K_{ai}=0.017\mu M$, for varying $K_{r_2s}$ and PGE2 concentrations  $0.01\mu{\rm M}, 0.1\mu{\rm M}, 1\mu{\rm M}, 10\mu{\rm M}$}
    \label{im:EP2_EP4_Kr2s_var}
\end{figure}

\begin{figure}[ht]
  \centering
    \includegraphics[height=220pt]{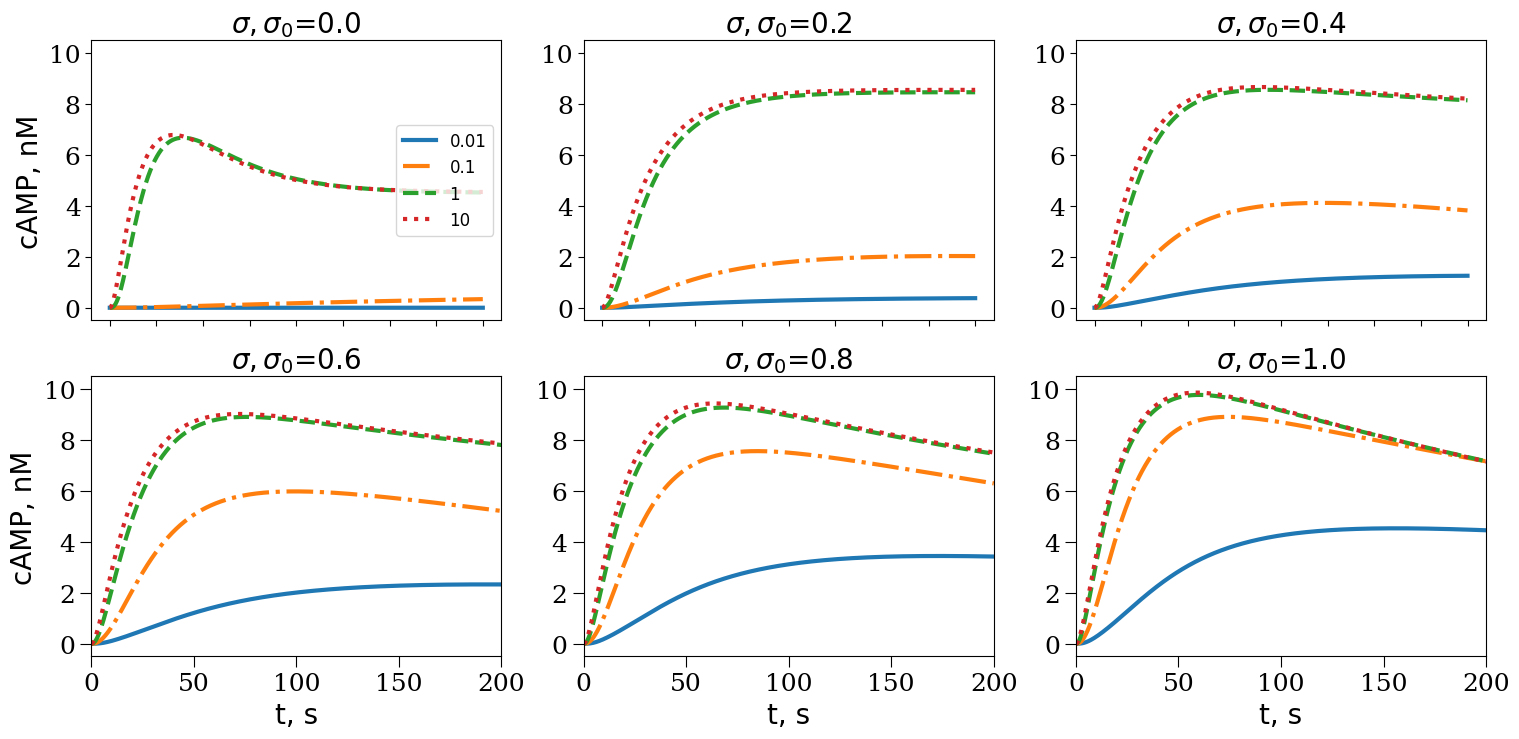}
    \caption{Simulated cAMP levels for the combined EP2-EP4 signaling pathway; model \eqref{eq:ligands_coupled}--\eqref{eq:EP2-EP4}  with parameter values in Tables~2--4, except $K_{ai}=0.017\mu M$, for varying $\sigma=\sigma_0$ and PGE2 concentrations  $0.01\mu{\rm M}, 0.1\mu{\rm M}, 1\mu{\rm M}, 10\mu{\rm M}$}
    \label{im:EP2_EP4_sigma_sigma_0_var}
\end{figure}

\end{document}